\documentclass[rpreprint,3p,10pt,sort&compress, colorlinks,linkcolor=blue, citecolor=blue,authoryear]{elsarticle}
\usepackage{amssymb}
\usepackage{amsmath,mathrsfs}
\usepackage{stfloats}
\usepackage{bm}
\usepackage{graphicx}
\usepackage{float}
\usepackage{caption}
\usepackage{sectsty}

\usepackage[toc,page]{appendix}
\usepackage[colorlinks,linkcolor=blue,anchorcolor=green,citecolor=blue]{hyperref}

\journal{IJES}

\begin{document}

\begin{frontmatter}
\title{Phase-field modeling of non-isothermal grain coalescence in the unconventional sintering techniques}
\author[els]{Yangyiwei Yang}

\author[els]{Min Yi\corref{cor1}}
\ead{yi@mfm.tu-darmstadt.de}

\author[els]{Bai-Xiang Xu\corref{cor1}}
\ead{xu@mfm.tu-darmstadt.de}

\author[rvt]{Long-Qing Chen}

\cortext[cor1]{Corresponding author}

\address[els]{Mechanics of Functional Materials Division, Institute of Materials Science, Technische Universit\"at Darmstadt, Darmstadt 64287, Germany}

\address[rvt]{Department of Materials Science and Engineering, The Pennsylvania State University, University Park, PA, 16802, USA}


\begin{abstract}
A thermodynamically consistent phase-field model is developed to study the non-isothermal grain coalescence during the sintering process, with a potential application to the simulation in unconventional sintering techniques, e.g. spark plasma sintering, field-assisted sintering, and selective laser sintering, where non-equilibrium and high temperature gradient exist. In the model, order parameters are adopted to represent the bulk and atmosphere/pore region, as well as the crystallographic orientations. Based on the entropy analysis, the temperature-dependent free energy density is developed, which includes contributions from the internal energy (induced by the change of temperature and order parameters) and the order parameter related configurational entropy. The temperature-dependent model parameters are determined by using the experimental data of surface and grain boundary energies and interface width. From laws of thermodynamics, the kinetics for the order parameters and the order-parameter-coupled heat transfer are derived. The model is numerically implemented by the finite element method. Grain coalescence from two identical particles shows that non-isothermal condition leads to the unsymmetric morphology and curved grain boundary due to the gradients of on-site surface and grain-boundary energies induced by the local temperature inhomogeneity. More simulations on the non-isothermal grain coalescence from two non-identical and multiple particles present the temporal evolution of grain shrinkage/growth, neck growth, and porosity, demonstrating the capability and versatility of the model. It is anticipated that the work could provide a contribution to the research community of unconventional sintering techniques that can be used to model the non-isothermal related microstructural features.

\end{abstract}


\end{frontmatter}


\section{Introduction}
As one of the most important technological processes in the ceramic and powder metallurgy industries, sintering is widely known as a complex process involving multiple physical processes which can be classified as but not limited to (a) the mass transportations: diffusions through surface, interface (grain boundaries), volume and atmosphere (vaporization and condensation), and viscosity or liquid flow; (b) the structure relaxation: the grain-boundary migration, deformation and rigid-body motions of the particles, and plastic flow. The collective effects of above processes result in a porous bulk with two distinct phenomena, i.e., the coarsening which leads to the elimination of the pores (or the decrease of the total surface area), and the grain growth which results in the reduction of the total grain-boundary energy \citep{1kang2004sintering,2olevsky1998theory,3german1996sintering,4wong1979models}. These two phenomena are also termed as "grain coalescence" altogether \citep{3german1996sintering,german2009liquid}. In the viewpoint of accelerating the optimum design of sintering process, a rational theory for the grain coalescence model, which considers the above-mentioned phenomena, and the associated numerical simulations are indispensible.

From the middle of last century, many efforts have been made to develop the theoretical models to investigate the evolution of such porous bulk and predict the processibility and quality of the target material. Regarding the coarsening, grain growth, and their interactions, the kinetic approach was usually applied to characterize the grain coalescence during sintering process \citep{1kang2004sintering,2olevsky1998theory,5kraft2004numerical,6ashby1974first,7beere1975second,8johnson1969new}. Several remarkable modeling attempts, such as the two-particle model \citep{9frenkel1945viscous,10kuczynski1949self}, the cylindrical pore channel model \citep{11coble1961sintering} and the spherical pore model \citep{12mackenzie1949phenomenological}, have been developed in terms of a phenomenological viewpoint of the grain coalescence with symbolic microstructures. But they were based on the rather simple arrangement (e.g. two-particle system) with highly idealized geometry. The microstructure evolution during sintering was then represented by various diagrams of the simple particle array, such as the temperature-density diagram (or the sintering diagram) and the grain-pore diagram (or the grain growth diagram) \citep{6ashby1974first,14kurtz1980microstructure,15carpay1977discontinuous,16yan1981microstructural}. Another approach focused on the thermodynamical quantities in the sintering and related them to the observerable features of the microstructure (which will be also featured in this work), e.g. the specific surface and interface (grain-boundary) energy and dihedral angle related by Young’s equation \citep{17german2010thermodynamics,18li2001compacting,19lange1989thermodynamics}. This approach, however, has the limitation due to issues such as the isothermal and equilibrium assumption, and the neglect of some structural details (e.g. the neck formation and the particle deformation through the mass transportation). Those reasons make the previous models incapable of dealing with the geometry/arrangement and conditions in the realistic multi-particle sintering systems, e.g. the powder bed, powder stark or pressed billet where the size, shape and arrangement of the particles might vary drastically.

Theoretical models combining numerical methods were thereby introduced to simulate the grain coalescence, since they require less assumption and approximation than the theoretical analysis and can interact closely with the experimental research. Featured methods include the Monte-Carlo method \citep{20akhtar1994monte,21tandon1999monte,22tikare2010numerical}, cellular automata \citep{23pimienta1992cellular}, molecular dynamics \citep{24zhu1996sintering,25raut1998sintering} and phase field method \citep{26chockalingam20162d,27wang2006computer,28kumar2010phase,29asp2006phase}. In recent decades, phase-field method has been widely used in the simulation of the microstructure evolution for various physical processes \citep{30boettinger2002phase,31chen2002phase,32moelans2008introduction}, in particular, the diffusion-based mass transportation \citep{29asp2006phase,33cahn1961spinodal,34gugenberger2008comparison} and the grain growth \citep{35fan1997effect,36chen1994computer,37fan1997diffusion} which are essential in the modeling of the grain coalescence. This method is able to model the complex spatial geometries without explicitly tracking the position of the surface and interface \citep{27wang2006computer}. It assumes the diffusive interface and can reproduce the behavior of the surface and grain boundary \citep{38cahn1958free}, such as the diffusive anisotropy due to the relaxation and amorphous around the surface and grain boundary. The interface anisotropic kinetics can be addressed by setting the mobility as tensor \citep{34gugenberger2008comparison}. The essence to model the grain coalescence using phase-field method is the choice of proper order parameters corresponding to different physical laws, and the development of thermodynamically consistent free energy of the sintering system. Also, the model should be able to describe the diffusion-based mass transportation and grain growth. Basically, two kinds of order parameters are used: the conserved order parameter governed by the Cahn-Hilliard equation which characterizes substance-pores \citep{38cahn1958free,40wang1993thermodynamically}, and the non-conserved parameter governed by the Allen-Cahn equation which characterizes grains \citep{39allen1979microscopic,36chen1994computer}. One of the earliest attempts to combine such efforts into a phase-field grain coalescence model was carried out by Wang et al., who utilized a conserved density field and a set of non-conserved orientation fields \citep{40wang1993thermodynamically,41kazaryan1999generalized}. This model is then applied to simulate the nanoparticle sintering \citep{26chockalingam20162d} as well as the coarsening of porous polycrystal structure in several materials \citep{42ahmed2013phase}. Based on this model, some featured details, such as the dihedral angle, neck growth and shrinkage, and the grain growth kinetics with porosity, have been captured and validated by experiments \citep{42ahmed2013phase,43ahmed2014phase,44millett2012phase,45biner2016pore}. Nevertheless, this model is established under the isothermal conditions, even though the mobilities can be obtained on different temperature following the Arrhenius equations. Thus, this model is limited to the conventional sintering in which the temperature is strictly controlled.

Recently, apart from the traditional sintering technique, there are increasing interests in the development of new manufacturing techniques for highly efficient and shape-controlled material production, such as the spark plasma sintering \citep{46guillon2014field,47munir2006effect,48anselmi2005fundamental,49yucheng2002study}, field-assisted sintering \citep{46guillon2014field,50rathel2009temperature,51vanmeensel2005modelling} and selective laser sintering \citep{52yap2015review,53kumar2003selective}. These new techniques often possess extraordinary features including local high-energy input, non-equilibrium, fast heating/cooling, high temperature gradient, etc. To improve the understanding of the sintering process in these new techniques, a phase-field model suitable for the non-isothermal grain coalescence cases is needed, which is, however, less investigated. Establishing such a model would face the challenges, such as the introduction of the temperature-dependent quantities and their interaction with the order parameters. Coupling with the thermal process (e.g. heat transfer) can be also challenging since most current theoretical interpretations on this aspect are based on assumptions. The difference between the time scale of the thermal process and grain coalescence should also be taken into account. In addition, a thermodynamically consistent model, whose model parameters have clear physical meaning and can be obtained from experimental results, is highly desired. The present work is focused on developing a non-isothermal phase-field model for the unconventional sintering techniques, as well as its finite-element numerical implementation. The model is derived in a thermodynamically consistent way and coupled with the temperature field. The parameterization is accomplished by using the experimentally measurable quantities. It is able to capture interesting phenomena which are not accessible to the conventional isothermal model. The purpose of the present work is to provide a novel contribution to the research community of unconventional sintering that can be used to model the non-isothermal phenomenon related thermal and microstructural features.

The paper is organized as follows. In Section \ref{pfmodel}, the phase-field model is derived in a thermodynamically consistent framework. The temperature-dependent free energy formulation of the model and the coupling between the phase fields and temperature are constructed thermodynamically. Allen--Cahn and Cahn--Hilliard equations are used to describe the kinetics of the order parameters. In Section \ref{numeric}, numerical implementation of the model is carried out by using finite element method in the framework of multiphysics object-oriented simulation (MOOSE) environment \citep{60tonks2012object}. In Section \ref{resultdis}, numerical examples are presented to show the basic feature and the ability of the model, including the benchmark of temperature-dependent surface and grain-boundary energies and dihedral angle and non-isothermal grain coalescence from two particles and particle assemblies. In particular, the connection of the model parameters to the physical quantities which can be experimentally measured is demonstrated. Section \ref{conclu} contains concluding remarks and future directions. Finally, the appendix summarizes the details on the derivation of the surface and grain-boundary energies formulation at equilibrium, the model parameters, the evolution equation for the internal energy density, and the residuals and tangent matrices for finite element implementation.

\section{Phase-field model formulation}\label{pfmodel}
\subsection{Thermodynamics}\label{thermo}
\begin{figure}[!b]
\centering
\includegraphics[width=11cm]{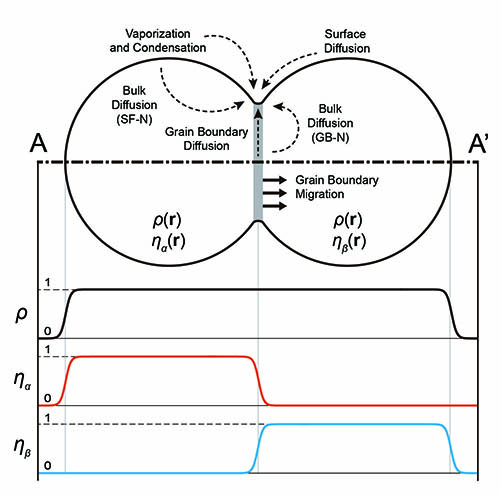}
\caption{Schematics for the phase-field interpretation of the sintering system by using the conserved order parameter (the fractional density field) $\rho$ and the multiple non-conserved order parameters (the orientation fields) $\{\eta_\alpha\}$. An illustration of order parameters ($\rho, \eta_\alpha, \eta_\beta$) profile across A-A' section is given. Featured physical phenomena, including different diffusion mechanisms and grain boundary migration, are also presented.}
\label{fig1}
\end{figure}

In this work, both conserved and non-conserved order parameters $\rho$ and $\{\eta_\alpha\}$ are utilized. Following \cite{27wang2006computer} and \cite{42ahmed2013phase}, the conserved phase-field parameter $\rho$ is the fractional density field. $\rho=1$ and $\rho=0$ represent the bulk and atmosphere/pore region, respectively. The non-conserved phase-field parameter $\eta_\alpha$ is the orientation fields which are set to distinguish particles/grains with different crystallographic orientations. It is anticipated that a general sintering system with multiple particles can have a spatial and temporal density field $\rho(\mathbf{r},t)$ in combination with the orientation field $\{\eta_\alpha\}=\{\eta_1(\mathbf{r},t),\eta_2(\mathbf{r},t),\dots \eta_N(\mathbf{r},t)\}$. The number of the orientation field parameters $N$ is not necessarily the same as the number of the grains.  In each grain, one of $\eta_\alpha$ is 1 and the others equal to zero (Fig. \ref{fig1}). Meanwhile, these grains simultaneously have $\rho=1$ including the grain boundaries (we assume the density variation across the grain boundary is negligible). When $\rho=0$, no grain presents. This profile of order parameters leads to the constraint $(1-\rho)+\sum_{\alpha} \eta_\alpha=1$. Their temporal evolution with the time indicate the changes of the surface and grain boundary and reveal the grain coalescence during sintering.

The kinetic equations of order parameters need the determination of the system free energy, which should be also temperature-dependent in the non-isothermal case. In this regard we start with the Legendre transform \citep{ruelle1999statistical}:
\begin{equation}
\mathscr{F}=\inf_{\mathscr{E}} \left[\mathscr{E}-T\mathscr{S}\right],
\label{eq1}
\end{equation}
where $\mathscr{F}$ represents the free energy, $\mathscr{E}$ the internal energy, $\mathscr{S}$ the entropy and $T$ the temperature. In a sintering system with multigrain and pores, the entropy $\mathscr{S}$ of a finite subdomain $\Omega$ within the system can be written in a functional form \citep{55penrose1990thermodynamically}: 
\begin{equation}
\mathscr{S}(e,\rho,\{\eta_\alpha\})=\int_\Omega \left[ s(e,\rho,\{\eta_\alpha\}) - \frac{1}{2}\kappa_\rho \left| \nabla \rho \right|^2 - \frac{1}{2}\kappa_\eta \sum_\alpha \left| \nabla \eta_\alpha \right|^2 \right] \text{d}\Omega .
\label{eq2}
\end{equation}
Here, the entropy density $s$ is only the function of the order parameters $\rho,\eta_\alpha$ and the internal energy density $e$. The positive constants $\kappa_\rho$ and $\kappa_\eta$ denote the contribution to the entropy density from the gradient of the order parameter according to the gradient thermodynamics \citep{38cahn1958free}. To relate this entropy functional to the free energy which is also a functional of the order parameters and temperature, we assume that the internal energy $\mathscr{E}$ is only the integration of the internal energy density $e(\rho,\{\eta_\alpha\})$ through the subdomain, then substitute Eq. (\ref{eq2}) into Eq. (\ref{eq1}) and obtain
\begin{equation}
\begin{split}
\mathscr{F}(T,\rho,\{\eta_\alpha\}) & =\inf_{\mathscr{E}} \left[\mathscr{E}(\rho,\{\eta_\alpha\}) - T \mathscr{S}(e,\rho,\{\eta_\alpha\}) \right] \\
& = \inf_{\mathscr{E}} \left[ \int_\Omega \left[ e(\rho,\{\eta_\alpha\}) - Ts(e,\rho,\{\eta_\alpha\}) + \frac{1}{2}T\kappa_\rho \left| \nabla \rho \right|^2 + \frac{1}{2}T\kappa_\eta \sum_\alpha \left| \nabla \eta_\alpha \right|^2 \right] \text{d}\Omega \right], \\
& =  \int_\Omega\left[\inf_{e}\left[ e(\rho,\{\eta_\alpha\}) - Ts(e,\rho,\{\eta_\alpha\})\right] + \frac{1}{2}T\kappa_\rho \left| \nabla \rho \right|^2 + \frac{1}{2}T\kappa_\eta \sum_\alpha \left| \nabla \eta_\alpha \right|^2\right] \text{d}\Omega .
\label{eq3}
\end{split}
\end{equation}
When the entropy density $s$ is lower bounded on $e$  at constant $\rho$ and $\{\eta_\alpha \}$, it is available to find a free energy density $f$ which is also lower bounded on $T$ at constant $\rho$  and $\{\eta_\alpha \}$  according to the Legendre transform, i.e.
\begin{equation}
f(T,\rho,\{\eta_\alpha\})=\inf_{e}\left[ e(\rho,\{\eta_\alpha\})-Ts(e,\rho,\{\eta_\alpha\}) \right].
\label{eq4}
\end{equation}
Then the free energy can be thereby formulated as
\begin{equation}
\mathscr{F}(T,\rho,\{\eta_\alpha\})=\int_\Omega \left[ f(T,\rho,\{\eta_\alpha\}) + \frac{1}{2} T \kappa_\rho \left| \nabla \rho \right|^2 + \frac{1}{2} T \kappa_\eta \sum_\alpha \left| \nabla \eta_\alpha \right|^2 \right] \text{d}\Omega .
\label{eq5}
\end{equation}
From Eq. (\ref{eq4}), one can obtain the following relation $\text{d}(f/T)=e\text{d}(1/T)$ \citep{55penrose1990thermodynamically}. Then an explicit formulation of $f$ can be obtained by integrating both sides with respect to $1/T$, which is
\begin{equation}
\frac{f(T,\rho,\{\eta_\alpha\})}{T}= \int e(\rho,\{\eta_\alpha\}) \text{d} \left(\frac{1}{T} \right).
\label{eq6}
\end{equation}
Assuming $e$ is formulated as
\begin{equation}
e(\rho,\{\eta_\alpha\})= e_\text{ht}(T)h(\rho,\{\eta_\alpha\}) + e_\text{pt}(\rho,\{\eta_\alpha\}),
\label{eq7}
\end{equation}
where $e_\text{ht}$ is the gain (or loss) of the internal energy density from the temperature change, and $e_\text{pt}$ is the spatial distribution of the internal energy density with respect to the order parameters $\rho$ and $\{\eta_\alpha\}$. $h(\rho,\{\eta_\alpha\})$ is a monotonic increasing function which varies smoothly with respect to order parameters and maps the heat to the regions with certain value of the order parameters, called the "interpolating function" \citep{30boettinger2002phase,32moelans2008introduction}. Substituting Eq. (\ref{eq7}) into Eq. (\ref{eq6}) and adding a temperature independent term $s_\text{cf}(\rho,\{\eta_\alpha\})$  after integration with respect to $1/T$, we obtain
\begin{equation}
f(T,\rho,\{\eta_\alpha\})= f_\text{ht}(T)h(\rho,\{\eta_\alpha\}) + e_\text{pt}(\rho,\{\eta_\alpha\}) - Ts_\text{cf}(\rho,\{\eta_\alpha\}),
\label{eq8}
\end{equation}
where
\begin{equation*}
f_\text{ht}(T)= T \int  e_\text{ht}(T) \text{d}\frac{1}{T}.
\end{equation*}
$s_\text{cf}$ is a term with the unit of entropy and only related to the order parameters, usually known as the configurational entropy \citep{55penrose1990thermodynamically,55penrose1990thermodynamically}. If the Legendre transform in Eq. (\ref{eq4}) is read in the inversion form as \citep{55penrose1990thermodynamically,ruelle1999statistical}
\begin{equation}
s(e,\rho,\{\eta_\alpha\}) = \inf_{e}\left[ \frac{e(\rho,\{\eta_\alpha\})-f(T,\rho,\{\eta_\alpha\})}{T} \right].
\label{eq9}
\end{equation}
It can be verified that $s_\text{cf}$ belongs to the entropy contribution by substituting Eqs. (\ref{eq7}) and (\ref{eq8}) into Eq. (\ref{eq9}), which yields
\begin{equation}
s(e,\rho,\{\eta_\alpha\}) = s_\text{ht}(T)h(\rho,\{\eta_\alpha\}) + s_\text{cf}
\label{eq10}
\end{equation}
with the thermal entropy
\begin{equation*}
s_\text{ht}(T)=\frac{e_\text{ht}-f_\text{ht}(T)}{T}=\int \frac{\text{d}e_\text{ht}}{T}.
\end{equation*}
Eqs. (\ref{eq8}) and (\ref{eq10}) are only a general formulation for the free energy density in the non-isothermal case. The detailed formulation of terms such as $e_\text{ht}$, $e_\text{pt}$ and $s_\text{cf}$ should be given according to the features of the grain coalescence, which will be sequentially discussed in the Subsection \ref{free}.

\subsection{Free energy density}\label{free}

\begin{figure}[!t]
\centering
\includegraphics[width=16cm]{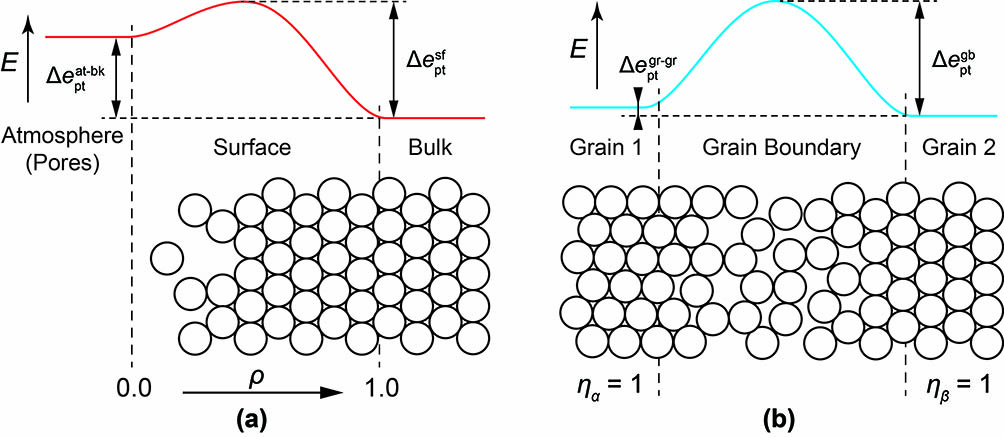}
\caption{Illustrations of the potential profile across (a) the surface and (b) the grain boundary at the initial temperature $T_0$. Here the $\Delta e_\text{pt}^\text{at-bk}$ and $\Delta e_\text{pt}^\text{gr-gr}$ are the potential differences between atmosphere/pores and bulk as well as two grains; $\Delta e_\text{pt}^\text{sf}$ and $\Delta e_\text{pt}^\text{gb}$ are the barrier heights on the surface and grain boundary, respectively.}
\label{fig2}
\end{figure}

Eq. (\ref{eq8}) implies that we can formulate the free energy density $f$ through the construction of the internal energy density $e$ of the system. It contains the spatial distribution term $e_\text{pt}$ inheriting from the potential term of the internal energy density, the heat-contribution term $f_\text{ht} h(\rho,\{\eta_\alpha\})$, and configuration term  $Ts_\text{cf}$. In most cases of the sintering, regions around surface and grain boundary always possess a higher energy than the bulk (grains) or the atmosphere/pore (Fig. \ref{fig2}). The system tends to eliminate those regions through physical processes like diffusion and grain boundary migration to reduce the total energy of the system \citep{1kang2004sintering,17german2010thermodynamics}. Setting the internal energy of the atmosphere/pores at an initial temperature $T_0$ of the system as the zero potential, we can simply give the formulation of the potential term as
\begin{equation}
e_\text{pt}(\rho,\{\eta_\alpha\}) = (1-\rho) \Delta e_\text{pt}^\text{at-bk} + \sum_\alpha \eta_\alpha \Delta e_\text{pt}^\text{gr-gr} + w_\text{pt}(\rho,\{\eta_\alpha\}),
\label{eq11}
\end{equation}
where $\Delta e_\text{pt}^\text{at-bk}$ is the potential difference between atmosphere/pores and bulk, and $\Delta e_\text{pt}^\text{gr-gr}$ the potential difference between two grains. Although these two terms contribute to the surface and grain-boundary energy, respectively, $\Delta e_\text{pt}^\text{at-bk}$ is usually treated as zero for the sintering of single chemical component, and   is difficult to be practically measured. Therefore, the only term can be formulated in Eq. (\ref{eq11}) is the so-called "multi-well" term $ w_\text{pt}(\rho,\{\eta_\alpha\})$. Here we formulate $ w_\text{pt}(\rho,\{\eta_\alpha\})$ with a Landau-type polynomial \citep{27wang2006computer,57landau1968statistical}
\begin{equation}
w_\text{pt}(\rho,\{\eta_\alpha\})=\underline{C}_\text{pt}\left[\rho^2(1-\rho)^2 \right] + \underline{D}_\text{pt}\left[\rho^2+6(1-\rho)\sum_\alpha\eta_\alpha^2 - 4(2-\rho)\sum_\alpha\eta_\alpha^3 + 3 \left(\sum_\alpha\eta_\alpha^2 \right)^2  \right],
\label{eq12}
\end{equation}
where $\underline{C}_\text{pt}$ and $\underline{D}_\text{pt}$ are model parameters which have to be determined for the specific material system. Eq. (\ref{eq12}) presents the local minima at $\rho=0$ and $\rho=1$, and a barrier height $\Delta e_\text{pt}^\text{sf}=(\underline{C}_\text{pt}+7\underline{D}_\text{pt})/16$ between minima. Around the surface $\rho$ and at most one of $\{\eta_\alpha \}$ smoothly varies from 0 to 1, or \textit{vice versa}. As the order parameter profile shown in Fig. \ref{fig1}, all $\{\eta_\alpha \}$ are forced to be zero when $\rho$ equals zero. On the other hand, $\rho$ remains 1 across the grain boundary, and $\{\eta_\alpha \}$ varies with different grains. Eq. (\ref{eq12}) also presents $N$ local minima for $N$ orientations at corresponding $\eta_\alpha=1$  and a barrier height $\Delta e_\text{pt}^\text{gb}=3\underline{D}_\text{pt}/4$  between every two minima. Therefore, $N+1$ minima, including ($\rho=0, \{\eta_1=0,\eta_2=0, \dots , \eta_N=0 \}$) for an atmosphere/pores state and ($\rho=1, \{\eta_1=1,\eta_2=0, \dots , \eta_N=0 \}$, ($\rho=1, \{\eta_1=0,\eta_2=1, \dots , \eta_N=0 \}$), \dots , ($\rho=1, \{\eta_1=0,\eta_2=0, \dots , \eta_N=1 \}$) for grain states with $N$ orientations, should hold at any temperature. Correspondingly, barriers $\Delta e_\text{pt}^\text{sf}$ and $\Delta e_\text{pt}^\text{gb}$ show relatively higher thermodynamic potential on the surface and grain boundaries, which is the origination of the driving force for the diffusion and the grain boundary migration, respectively. Therefore, these barriers should directly relate to the derivation of the surface and grain-boundary energy which we elaborated in \ref{appendixa}.

Although $w_\text{pt}(\rho,\{\eta_\alpha\})$ shows the same zero minima which are already sufficient for the modeling of the isothermal case \citep{27wang2006computer,42ahmed2013phase}, further interpretation for the temperature-dependent terms, including the heat contribution term $f_\text{ht}h(\rho,\{\eta_\alpha\})$  as well as the configuration entropy term $Ts_\text{cf}$, are required for the non-isothermal cases. If no phase transition exists within the range of the temperature gradient, a.k.a. the solid-state sintering cases, we can formulate the heat term of the internal energy density within bulk as (note that we have set the internal energy of the atmosphere/pores at an initial temperature $T_0$ of the system as the zero potential)
\begin{equation}
e_\text{ht}(T)=(c_\text{bk}m_\text{bk}-c_\text{at}m_\text{at})(T-T_0) = c_r(T-T_0),
\label{eq13}
\end{equation}
where $c_\text{bk}$, $c_\text{at}$, $m_\text{bk}$ and $m_\text{at}$ are the specific heat and the real density of the bulk and the atmosphere/pores respectively, which can be represented by a relative specific heat $c_\text{r}$. Contribution from heat to the free energy density $f_\text{hc}$ can be thereby transformed from $e_\text{ht}$ accordingly. Assuming each grain has the same heat transfer under a certain temperature difference, we only need the interpolating function $h(\rho,\{\eta_\alpha\})$ to map the heat term onto the bulk region. Here we formulate $h(\rho,\{\eta_\alpha\})$ as
\begin{equation}
h(\rho,\{\eta_\alpha\})=\underline{A}\rho + \underline{B}\sum_\alpha \eta_\alpha,
\label{eq14}
\end{equation}
where $\underline{A}$ and $\underline{B}$ are also model parameters and should satisfy $\underline{A}+\underline{B}=1$. There is always $\rho$ and at most one $\eta_\alpha$ varies from zero to one across the surface as shown in Fig. \ref{fig1}. Eq. (\ref{eq14}) can thereby vary smoothly from zero to one across the surface and remain one within the bulk. Another feature of Eq. (\ref{eq14}) is that the model parameters can be directly related to the surface and grain-boundary energies as elaborated in \ref{appendixa}. As for the configuration entropy $s_\text{cf}$, it has a similar profile as the potential in Fig. \ref{fig2}. The minimum entropy can be found in atmosphere/pores and multiple grains and the maximum on the surface and grain boundaries. Here by adopting $w_\text{pt}(\rho,\{\eta_\alpha\})$ to  $s_\text{cf}$ with two model parameters $\underline{C}_\text{cf}$ and $\underline{D}_\text{cf}$, we have
\begin{equation}
s_\text{cf}(\rho,\{\eta_\alpha\})=\underline{C}_\text{cf}\left[\rho^2(1-\rho)^2 \right] + \underline{D}_\text{cf}\left[\rho^2+6(1-\rho)\sum_\alpha\eta_\alpha^2 - 4(2-\rho)\sum_\alpha\eta_\alpha^3 + 3\left(\sum_\alpha\eta_\alpha^2 \right)^2 \right].
\label{eq15}
\end{equation}
Combining Eqs. (\ref{eq12})--(\ref{eq15}), formulation of the free energy density eventually shows as following (select the maximum temperature within the system as the initial temperature $T_0$):
\begin{equation}
\begin{split}
f(T,\rho,\{\eta_\alpha\})=&f_\text{ht}(T)\left(\underline{A}\rho + \underline{B}\sum_\alpha\eta_\alpha\right) +
\underline{C}\left[\rho^2(1-\rho)^2 \right] + \\
&\underline{D}\left[\rho^2+6(1-\rho)\sum_\alpha\eta_\alpha^2 - 4(2-\rho)\sum_\alpha\eta_\alpha^3 + 3\left(\sum_\alpha\eta_\alpha^2 \right)^2 \right],
\label{eq16}
\end{split}
\end{equation}
where
\begin{equation*}
\begin{split}
& f_\text{ht}(T)=T\int e_\text{ht}(T)\text{d}\left(\frac{1}{T}\right)=-c_\text{r}T\text{ln}\frac{T}{T_0}+c_\text{r}(T-T_0) , \\
& \underline{C}=\underline{C}_\text{pt} - \underline{C}_\text{cf}(T-T_0) , \\
& \underline{D}=\underline{D}_\text{pt} - \underline{D}_\text{cf}(T-T_0).
\end{split}
\end{equation*}
At equilibrium, the eight model parameters ($\underline{A},\underline{B},\underline{C}_\text{pt}, \underline{C}_\text{cf}, \underline{D}_\text{pt}, \underline{D}_\text{cf},~\kappa_\eta$ and $\kappa_\rho$) should have a dependency through Eq. (\ref{eqA7}). Thus we have
\begin{equation*}  
\underline{A}=\frac{\kappa_\rho}{\kappa_\rho+\kappa_\eta}, ~ \underline{B}=\frac{\kappa_\eta}{\kappa_\rho+\kappa_\eta}, ~
\frac{\underline{C}_\text{pt}+\underline{D}_\text{pt}}{\kappa_\rho}=\frac{6\underline{D}_\text{pt}}{\kappa_\eta}, ~
\frac{\underline{C}_\text{cf}+\underline{D}_\text{cf}}{\kappa_\rho}=\frac{6\underline{D}_\text{cf}}{\kappa_\eta}.
\end{equation*}

\begin{figure}[!t]
\centering
\includegraphics[width=16cm]{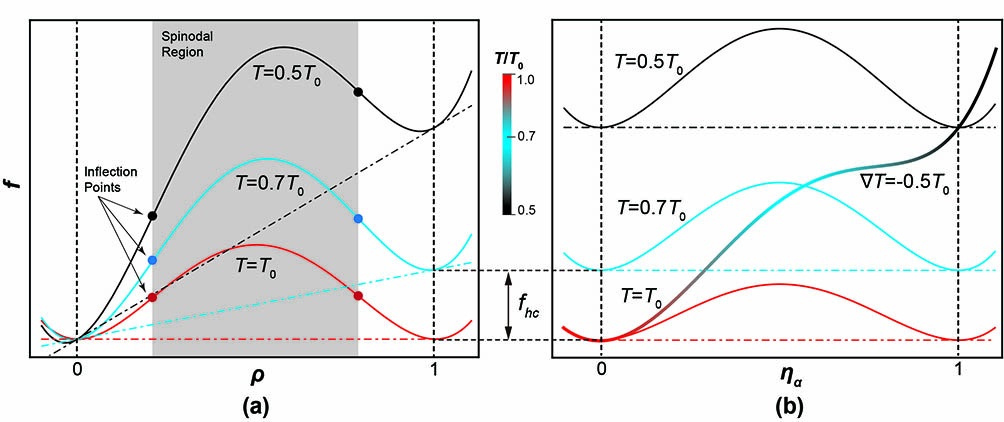}
\caption{Scheme diagram of the profile of the free energy density across (a) the surface and (b) the grain boundary at various temperature. The thick line with gradient color in (b) shows the profile when a temperature gradient $\nabla T$ across the grain boundary.}
\label{fig3}
\end{figure}

These model parameters can be obtained from the experimentally measured temperature-dependent surface and grain-boundary energies and the width of the grain boundary. The derivation of the dependency and determination of these quantities is given in \ref{appendixa}.

Fig. \ref{fig3}a plots the temperature-dependent profile of the free energy density $f$ of the sintering system across the surface where $\rho\in[0,~1]$. According to the Maxwell relations of the thermodynamic quantities $\partial\mathscr{F}/\partial T=-\mathscr{S}$. Since entropy $\mathscr{S}$ is positively defined, it can be shown that free energy decreases with the increase of $T$. We can clearly see that when temperature decreases from initial $T_0$ to 0.5$T_0$, the point representing the bulk state of the $f$-curve will be tiled up to the value $f_\text{ht}$ because the relative specific heat $c_\text{r}$ defined in Eq. (\ref{eq13}) is always positive in most cases. Meanwhile, the barrier height between two local minima is lifted, showing the contribution of the configuration term. Similar to the profile of the free energy density in the spinodal decomposition \citep{33cahn1961spinodal}, points within the spinodal region becomes thermodynamically unstable and spontaneously "slide" to one of the two tangent points of the $f$-curve and the tangent line $f_\text{ht}\rho$, which should be always posited on $\rho=0$ and $\rho=1$ to show the coherence with the mass conservation. This process is governed by the diffusion.

Fig. \ref{fig3}b illustrates the temperature-dependent profile of the free energy density $f$ of the sintering system across a grain boundary where $\rho=1$ and $\eta_\alpha,~\eta_\beta \in [0,~1]$. Due to the constraint $(1-\rho)+\sum_{\alpha} \eta_\alpha=1$, we can simply set $\eta_\beta=1-\eta_\alpha$. It shows that at each identical temperature the $f$-curve has the same local minima, indicating that thermodynamic stability of grains equally maintains. The contribution of the configuration term can be also seen from the lift of the barrier height from $T_0$ to $0.5T_0$. However, such symetricsymmetric multi-well will be tilted in the existence of the temperature gradient, and thus make the grains in the hot side more preferable thermodynamically \citep{58tonks2014demonstrating,59gottstein2009grain}.

\subsection{Kinetic equations}\label{kinetic}
Here we present the derivation of the kinetics for the conserved and non-conserved order parameters as well as heat transfer, which is based on the work by \cite{33cahn1961spinodal, 39allen1979microscopic, 55penrose1990thermodynamically,75penrose1993relation}, \cite{27wang2006computer} and \cite{55penrose1990thermodynamically} where entropy density functional is fundamentally formulated as shown in Eq. (\ref{eq2}) where the internal energy density $e$ and entropy density $s$ are related by
\begin{equation}
\text{d}e=T\text{d}s+\frac{\partial e}{\partial \rho }\text{d}\rho +\sum\limits_{\alpha }{\frac{\partial e}{\partial {{\eta }_{\alpha }}}\text{d}{{\eta }_{\alpha }}}.
\label{eqA20}
\end{equation}
According to the first law of the thermodynamics, the internal energy change of the certain subdomain and the thermal flux ${{\mathbf{j}}_{e}}$ on the close surface of the subdomain should obey
\begin{equation}
\int\limits_{\Omega }{\dot{e}\left( \rho ,\left\{ {{\eta }_{\alpha }} \right\} \right)\text{d}\Omega }+\int\limits_{\Gamma }{{{\mathbf{j}}_{e}}\cdot \mathbf{n}\text{d}\Gamma }=0,
\label{eqA21}
\end{equation}
where $\mathbf{n}$ is the normal vector of the surface $\Gamma $. Using the Gaussian theorem, Eq. (\ref{eqA21}) can be also written as
\begin{equation}
\dot{e}\left( \rho ,\left\{ {{\eta }_{\alpha }} \right\} \right)+\nabla \cdot {{\mathbf{j}}_{e}}=0.
\label{eqA22}
\end{equation}

Since the order parameter $\rho $, indicating the fractional density field, is conserved by the mass conservation, we can also take the following relation under consideration
\begin{equation}
\dot{\rho }+\nabla \cdot {{\mathbf{j}}_{\rho }}=0.
\label{eqA25}
\end{equation}

The time derivative of $\mathscr{S}\left( e,\rho ,\left\{ {{\eta }_{\alpha }} \right\} \right)$ can be given as
\begin{equation}
\begin{aligned}
\dot{\mathscr{S}}\left( e,\rho ,\left\{ {{\eta }_{\alpha }} \right\} \right)  =&\int\limits_{\Omega }{\left[ \frac{\partial s}{\partial e}\dot{e}+\frac{\partial s}{\partial \rho }\dot{\rho }\text{+}\sum\limits_{\alpha }{\frac{\partial s}{\partial {{\eta }_{\alpha }}}{{{\dot{\eta }}}_{\alpha }}}-{{\kappa }_{\rho }}\nabla \rho \cdot \nabla \dot{\rho }-\sum\limits_{\alpha }{{{\kappa }_{\eta }}\nabla {{\eta }_{\alpha }}\cdot \nabla {{{\dot{\eta }}}_{\alpha }}} \right]\text{d}\Omega } \\ 
=&\int\limits_{\Omega }{\left[ \frac{\partial s}{\partial e}\dot{e}+\left( \frac{\partial s}{\partial \rho }\text{+}{{\kappa }_{\rho }}{{\nabla }^{\text{2}}}\rho  \right)\dot{\rho }\text{+}\sum\limits_{\alpha }{\left( \frac{\partial s}{\partial {{\eta }_{\alpha }}}\text{+}{{\kappa }_{\eta }}{{\nabla }^{\text{2}}}{{\eta }_{\alpha }} \right){{{\dot{\eta }}}_{\alpha }}} \right]\text{d}\Omega } \\ 
&-\int\limits_{\Gamma }{\left[ {{\kappa }_{\rho }}\left( \dot{\rho }\nabla \rho  \right)+\sum\limits_{\alpha }{{{\kappa }_{\eta }}\left( {{{\dot{\eta }}}_{\alpha }}\nabla {{\eta }_{\alpha }} \right)} \right]\cdot \mathbf{n}\text{d}\Gamma }, \\ 
\end{aligned}
\label{eqA23}
\end{equation}
in which the integration term on the surface $\Gamma$ is usually neglected according to \cite{38cahn1958free}. Then according to the second law of thermodynamics, the total production of the entropy should always be non-negative \citep{bi1998phase,55penrose1990thermodynamically}. As for the subdomain $\Omega $ and its close surface $\Gamma $, the production rate of the entropy should formulate as
\begin{equation}
\mathscr{\dot{S}}+\int\limits_{\Gamma }{{{\mathbf{j}}_s^e}\cdot \mathbf{n}\text{d}\Gamma } +\int\limits_{\Gamma }{{{\mathbf{j}}_s^\rho}\cdot \mathbf{n}\text{d}\Gamma } \ge 0,
\label{eqA24}
\end{equation}
In Eq. (\ref{eqA24}), ${{\mathbf{j}}_{s}^e}$ and ${{\mathbf{j}}_{s}^\rho}$ represent the entropy flux from ${{\mathbf{j}}_{e}}$ and ${{\mathbf{j}}_{\rho}}$, respectively, i.e.
\begin{equation}
\mathbf{j}_{s}^e=\frac{\partial s}{\partial e}{{\mathbf{j}}_{e}} \quad \text{and}
\quad {{\mathbf{j}}_{s}^\rho}=\frac{\partial s}{\partial \rho} {{\mathbf{j}}_{\rho}}
\label{eqA24-1}
\end{equation}

Substituting Eqs. (\ref{eqA22})--(\ref{eqA23}) and (\ref{eqA24-1}) into Eq. (\ref{eqA24}) then yields
\begin{equation}
\int\limits_{\Omega }{\left[ {{\mathbf{j}}_{e}}\cdot \nabla \frac{\partial s}{\partial e}+{{\mathbf{j}}_{\rho }}\cdot \nabla \left( \frac{\partial s}{\partial \rho }\text{+}{{\kappa }_{\rho }}{{\nabla }^{\text{2}}}\rho  \right)\text{+}\sum\limits_{\alpha }{\left( \frac{\partial s}{\partial {{\eta }_{\alpha }}}\text{+}{{\kappa }_{\eta }}{{\nabla }^{\text{2}}}{{\eta }_{\alpha }} \right){{{\dot{\eta }}}_{\alpha }}} \right]\text{d}\Omega }\ge 0.
\label{eqA26}
\end{equation}
The non-negative entropy production can be thereby guaranteed by taking the following relations, 
\begin{subequations}
	\begin{equation}
	{{\mathbf{j}}_{\rho }}=\hat{\mathbf{M}}\nabla \left( \frac{\partial s}{\partial \rho }\text{+}{{\kappa }_{\rho }}{{\nabla }^{\text{2}}}\rho  \right)=\hat{\mathbf{M}}\nabla \left(  \frac{\delta\mathscr{S}}{\delta\rho }  \right),
	\label{eqa27a}
	\end{equation}
	\begin{equation}
	{{\dot{\eta }}_{\alpha }}=\hat{\mathbf{L}}\left( \frac{\partial s}{\partial {{\eta }_{\alpha }}}\text{+}{{\kappa }_{\eta }}{{\nabla }^{\text{2}}}{{\eta }_{\alpha }} \right)=\hat{\mathbf{L}}\left( \frac{\delta\mathscr{S}}{\delta\eta_{\alpha} } \right),
	\label{eqa27b}
	\end{equation}
	\begin{equation}
	{{\mathbf{j}}_{e}}=\hat{\mathbf{K}}\nabla \frac{\partial s}{\partial e}=\hat{\mathbf{K}}\nabla \frac{\delta\mathscr{S}}{\delta e },
	\label{eqa27c}
	\end{equation}
\end{subequations}
where $\hat{\mathbf{L}}$, $\hat{\mathbf{M}}$ and $\hat{\mathbf{K}}$ are all tensors. According to Legendre transform in Eq. (\ref{eq1}) \citep{28kumar2010phase}, we have
\begin{equation*}
\frac{\delta\mathscr{S}}{\delta\rho }=-\frac{1}{T}\frac{\delta\mathscr{F}}{\delta\rho } \quad \text{and} \quad
\frac{\delta\mathscr{S}}{\delta{{\eta }_{\alpha }}}=-\frac{1}{T}\frac{\delta\mathscr{F}}{\delta{{\eta }_{\alpha }}}.
\end{equation*}
If we adopt the tensor $\mathbf{M}$ with the form  $\mathbf{M}=\hat{\mathbf{M}} /T$, Eq. (\ref{eqa27a}) thereby results in the form of the Cahn-Hilliard equation \citep{33cahn1961spinodal} after substituting Eq. (\ref{eqA25}):
\begin{equation}
\dot{\rho}(\mathbf{r},t)=\nabla \cdot \left( \mathbf{M} \cdot \nabla \frac{\delta \mathscr{F}}{\delta \rho} \right) =
\nabla \cdot \left[ \mathbf{M} \cdot \nabla \left(\frac{\partial f}{\partial\rho}-T\kappa_\rho\nabla^2\rho \right) \right],
\label{eq17}
\end{equation}
where the tensor $\mathbf{M}$ represents the modified diffusive mobility which is formulated with the diffusivity tensor $\mathbf{D}$ as $\mathbf{M}=\mathbf{D}/2(\underline{C}+\underline{D})$ \citep{44millett2012phase,60tonks2012object}. Since there are various diffusion mechanisms during the grain coalescence (Fig. \ref{fig1}), the tensor $\mathbf{D}$ can be formulated by considering different diffusion routes reported in \citep{27wang2006computer,29asp2006phase,34gugenberger2008comparison,42ahmed2013phase}. Then we have
\begin{equation}
\mathbf{D}=\mathbf{D}_\text{bk}+\mathbf{D}_\text{vp}+\mathbf{D}_\text{sf}+\mathbf{D}_\text{gb},
\label{eq18}
\end{equation}
with
\begin{equation}
\begin{split}
& \mathbf{D}_\text{bk}=D_\text{bk}^\text{eff}\left[\rho^3\left(10-15\rho+6\rho^2\right)\right]\mathbf{I}, \\
& \mathbf{D}_\text{vp}=D_\text{vp}^\text{eff}\left[1-\rho^3\left(10-15\rho+6\rho^2\right)\right]\mathbf{I}, \\
& \mathbf{D}_\text{sf}=D_\text{sf}^\text{eff}\left[16\rho^2(1-\rho)^2\right]\mathbf{T}_\text{sf}, \\
& \mathbf{D}_\text{gb}=D_\text{gb}^\text{eff}\left[16\sum_{\alpha\neq\beta}\eta_\alpha^2\eta_\beta^2\right]\mathbf{T}_\text{gb}.
\end{split}
\label{eq19}
\end{equation}
$\mathbf{D}_\text{bk}$, $\mathbf{D}_\text{vp}$, $\mathbf{D}_\text{sf}$, and $\mathbf{D}_\text{gb}$ represents the diffusivity through bulk, vapor, surface, and grain boundaries, respectively. They can only have its effective quantities $D_\text{bk}^\text{eff}$, $D_\text{vp}^\text{eff}$, $D_\text{sf}^\text{eff}$, and $D_\text{gb}^\text{eff}$ in the corresponding region which can be temperature-dependent (e.g. obeying Arrhenius equations). $\mathbf{I}$ is the identity tensor while $\mathbf{T}_\text{sf}$ and $\mathbf{T}_\text{gb}$ are the projection tensor for surface and grain boundary which projects the effective value onto the surface and grain boundaries, respectively.

Similarly, if we adopt the tensor $\mathbf{L}$ with the form  $\mathbf{L}=\hat{\mathbf{L}} /T$, Eq. (\ref{eqa27b}) results in the form of the Allen-Cahn equation \citep{39allen1979microscopic}:
\begin{equation}
\dot{\eta}_\alpha(\mathbf{r},t)=-\mathbf{L} \frac{\delta \mathscr{F}}{\delta \eta_\alpha} =
-\mathbf{L} \left(\frac{\partial f}{\partial\eta_\alpha}-T\kappa_\eta\nabla^2\eta_\alpha \right),
\label{eq20}
\end{equation}
where tensor $\mathbf{L}$ is usually formulated as the scalar $L$ representing the grain-boundary mobility, a constant at each temperature when isotropic grain boundary migration is assumed. According to the quantitative analysis by \cite{61moelans2008quantitative}, $L$ can be explicitly formulated by using the grain-boundary mobility $G_\text{gb}^\text{eff}$, grain-boundary energy $\gamma_\text{gb}$ and the gradient model parameter $\kappa_\eta$ as
\begin{equation}
L=\frac{G_\text{gb}^\text{eff}\gamma_\text{gb}}{T\kappa_\eta}.
\label{eq21}
\end{equation} 
It should be noted that the gradient model parameter of Eq. (\ref{eq21}) has been modified from the original equation in Ref. \cite{61moelans2008quantitative} to $T\kappa_\eta$ for the physical coherence.

In order to derive the heat transfer equation, we start with the Eq. (\ref{eqa27c}) after substituting Eq. (\ref{eqA22}), which reads as
\begin{equation}
\dot{e}(\rho,~\{\eta_\alpha\})=-\nabla\cdot\left(\hat{\mathbf{K}}\cdot\nabla\frac{\delta\mathscr{S}}{\delta e}\right)
=-\nabla\cdot\left(\hat{\mathbf{K}}\cdot\nabla\frac{1}{T}\right),
\label{eq22}
\end{equation}
We then adopt $\hat{\mathbf{K}}=\mathbf{k}T^2$ where $\mathbf{k}$ is the thermal conductivity tensor by expanding the left-hand side. Then Eq. (\ref{eq22}) can be reformulated as
\begin{equation}
c_\text{r}\dot{T}(\mathbf{r},t) + \frac{\partial e}{\partial \rho} \dot{\rho}(\mathbf{r},t) + \sum_\alpha \frac{\partial e}{\partial \eta_\alpha} \dot{\eta}_\alpha(\mathbf{r},t) = - \nabla\cdot \left(\mathbf{k}\cdot\nabla T\right).
\label{eq23}
\end{equation}
Spatial distribution of $\mathbf{k}$ within the system is treated similarly as the diffusivity in Eq. (\ref{eq18}), i.e. 
\begin{equation}
\mathbf{k}=\mathbf{k}_\text{bk}+\mathbf{k}_\text{at}+\mathbf{k}_\text{sf}+\mathbf{k}_\text{gb},
\label{eq24}
\end{equation}
with
\begin{equation}
\begin{split}
& \mathbf{k}_\text{bk}=k_\text{bk}^\text{eff}\left[\rho^3\left(10-15\rho+6\rho^2\right)\right]\mathbf{I}, \\
& \mathbf{k}_\text{at}=k_\text{at}^\text{eff}\left[1-\rho^3\left(10-15\rho+6\rho^2\right)\right]\mathbf{I}, \\
& \mathbf{k}_\text{sf}=k_\text{sf}^\text{eff}\left[16\rho^2(1-\rho)^2\right]\mathbf{T}_\text{sf}, \\
& \mathbf{k}_\text{gb}=k_\text{gb}^\text{eff}\left[16\sum_{\alpha\neq\beta}\eta_\alpha^2\eta_\beta^2\right]\mathbf{T}_\text{gb}.
\end{split}
\label{eq25}
\end{equation}
$\mathbf{k}_\text{bk}$, $\mathbf{k}_\text{vp}$, $\mathbf{k}_\text{sf}$, and $\mathbf{k}_\text{gb}$ represents the thermal conductivity of bulk, atmosphere, surface, and grain boundaries, respectively. They can only have its effective quantities $k_\text{bk}^\text{eff}$, $k_\text{vp}^\text{eff}$, $k_\text{sf}^\text{eff}$, and $k_\text{gb}^\text{eff}$ in the corresponding region, which can be also temperature-dependent. Eq. (\ref{eq23}) gives the equation that not only governs the evolution of the local temperature, but also couples with the evolution of the bulks and grains. It demonstrates how the local temperature change interacts with grain coalescence. It is also easy to see that when the scale of the system is much larger than the size of the grain, i.e., the evolution of the bulk and the grains can be ignored, Eq. (\ref{eq23}) can be degraded to the conventional Fourier’s equation for heat transfer.

\section{Numerical implementation}\label{numeric}
\subsection{Normalization}
The dimensionless form is obtained by normalizing with respect to a set of reference quantities, including the reference (initial) temperature $T_0$, the reference energy density $\underline{C}_\text{pt}^{T_0}$ which is the corresponding model parameter obtained at the reference temperature $T_0$, the reference length scale $\lambda=\sqrt{\kappa_\rho^{T_0}T_0/C_\text{pt}^{T_0}}$, and the time scale $\tau=1 / L_r C_\text{pt}^{T_0}$. $\kappa_\rho^{T_0}$ is the gradient model parameter at a reference temperature $T_0$. $L_r$ is the properly-chosen reference grain-boundary mobility. Other quantities are given in Table \ref{tab1}. Spatial and time derivatives are also normalized with respect to $\tau$ and $\lambda$, respectively. 

\newcommand{\tabincell}[2]{\begin{tabular}{@{}#1@{}}#2\end{tabular}}
\renewcommand{\arraystretch}{1.5}
\begin{table} \small
\centering 
\caption{The dimensionless form of the quantities involved in this model.}
\begin{tabular}{cccccc}
\hline
 & Symbols & Normalization  &  &  Symbols & Normalization \\ \hline
\tabincell{c}{Physical \\ quantities} & 
\tabincell{c}{  $c$ \\ $k$ \\ $L$ \\ $M_{ij}$ } &  
\tabincell{c}{ $\widetilde{c}=cT_0/\underline{C}_\text{pt}^{T_0}$ \\ $\widetilde{k}=k \tau T_0 /(\underline{C}_\text{pt}^{T_0}\lambda^2)$ \\ $\widetilde{L}=L \tau \underline{C}_\text{pt}^{T_0}$  \\ $\widetilde{M}_{ij}=M_{ij} \tau \underline{C}_\text{pt}^{T_0} / \lambda^2$ } & 
\tabincell{c}{Model \\ parameters} & 
\tabincell{c}{$\underline{A}$ \\ $\underline{B}$ \\ $\underline{C}_\text{pt}$ \\ $\underline{D}_\text{pt}$ \\ $\underline{C}_\text{cf}$ \\ $\underline{D}_\text{cf}$ \\ $\kappa_\rho$ \\ $\kappa_\eta$ }  & 
\tabincell{c}{$\widetilde{\underline{A}}=\underline{A}$ \\ $\widetilde{\underline{B}}=\underline{B}$ \\ $\widetilde{\underline{C}}_\text{pt}=\underline{C}_\text{pt} / \underline{C}_\text{pt}^{T_0}$ \\ $\widetilde{\underline{D}}_\text{pt}=\underline{D}_\text{pt} / \underline{D}_\text{pt}^{T_0}$ \\ $\widetilde{\underline{C}}_\text{cf}=\underline{C}_\text{cf} T_0 / \underline{C}_\text{pt}^{T_0}$ \\ $\widetilde{\underline{D}}_\text{cf}=\underline{D}_\text{cf} T_0 / \underline{C}_\text{pt}^{T_0}$ \\ $\widetilde{\kappa}_\rho=\kappa_\rho T_0 / (\underline{C}_\text{pt}^{T_0}\lambda^2)$ \\ $\widetilde{\kappa}_\eta=\kappa_\eta T_0 / \underline{C}_\text{pt}^{T_0}\lambda^2)$} \\
    \hline
\end{tabular}
\label{tab1}
\end{table}

\subsection{Finite element implementation}
The model was numerically implemented within the MOOSE framework by finite element method (FEM) \citep{60tonks2012object}. The whole examples were performed on a two dimensional 100$\times$100 initial mesh with 9-node quadrilateral element, although the model itself is capable for three-dimensional cases. The Cahn-Hilliard equation in Eq. (\ref{eq17}), which is a 4th order differential equation, was solved by splitting it into two 2nd order differential equations by introducing an additional coupling field $\mu$ \citep{62elliott1989second,63zhao2015isogeometric,64balay1997efficient}. In this way, Eq. (\ref{eq17}) can be rewritten as
\begin{equation} 
 \dot{\rho}=\nabla \cdot (\mathbf{M}\cdot \nabla \mu) .
\label{eq26}
\end{equation}

\begin{equation} 
 \mu=\frac{\partial f}{\partial \rho} - T \kappa_\rho \nabla^2 \rho. 
\label{eq26-2}
\end{equation}

Then, weak forms of the Eqs. (\ref{eq20}), (\ref{eq22}), (\ref{eq26}), and (\ref{eq26-2}) are obtained by introducing corresponding trial functions $\psi_\mu$, $\psi_\rho$, $\psi_{\eta_\alpha}$ and $\psi_T$, respectively, and integrating by parts over the subdomain $\Omega$. For simplicity, the index notation of the tensor and its derivatives are used here. Then the weak forms read as
\begin{equation}
\begin{split}
& \int_\Omega\dot{\rho}\psi_\mu\text{d}\Omega=-\int_\Omega M_{ij}\mu_{,j}\psi_{\mu,i}\text{d}\Omega + \int_\Gamma M_{ij}\mu_{,j}\psi_\mu n_i \text{d}\Gamma, \\
& \int_\Omega\mu\psi_\rho \text{d}\Omega = \int_\Omega \frac{\partial f}{\partial \rho}\psi_\rho \text{d}\Omega + \int_\Omega \kappa_\rho T \rho_{,i} \psi_{\rho,i} \text{d}\Omega - \int_\Gamma \kappa_\rho T \rho_{,i} \psi_\rho n_i \text{d}\Gamma, \\
& \int_\Omega\dot{\eta}_\alpha\psi_\eta\text{d}\Omega= -L\int_\Omega\frac{\partial f}{\partial \eta_\alpha}\psi_\eta \text{d}\Omega - L\int_\Omega\kappa_\eta T \eta_{\alpha,i}\psi_{\eta,i}\text{d}\Omega + L\int_\Gamma\kappa_\eta T \eta_{\alpha,i}\psi_\eta n_i \text{d}\Gamma, \\
& \int_\Omega \frac{\partial e}{\partial T} \dot{T} \psi_T \text{d}\Omega + \int_\Omega\frac{\partial e}{\partial \rho} \dot{\rho} \psi_T \text{d}\Omega + \sum_\alpha \int_\Omega\frac{\partial e}{\partial \eta_\alpha} \dot{\eta}_\alpha\psi_T \text{d}\Omega =\int_\Omega k T_{,i}\psi_{T,i}\text{d}\Omega - \int_\Gamma k T_{,i}\psi_T n_i \text{d}\Gamma,
\end{split}
\label{eq27}
\end{equation}
where $n_i$ is the normal vector to the boundary $\Gamma$ of the subdomain. To solve the time-dependent PDEs, transient solver based on backward Euler algorithm has been employed. adoptive meshing and time stepping schemes are used to reduce the computation costs. The constraint of the order parameters is achieved using penalty functions. More details about the FEM implementation, such as residuals and iteration matrix, are shown in \ref{appendixc}.

\section{Results and discussion}\label{resultdis}
\subsection{Temperature-dependent surface and grain-boundary energies}\label{gbenergy}
The surface energy $\gamma_\text{sf}$ and grain-boundary energy $\gamma_\text{gb}$ can be integrated along the normal direction $\mathbf{r}$ of the surface and grain boundary by the Cahn's approach, i.e.
\begin{equation}
\begin{split}
& \gamma_\text{sf}=\int_{-\infty}^\infty \left[f(T,\rho,\{\eta_\alpha,\eta_\beta=0\}) + \frac{1}{2}T\kappa_\rho \left| \nabla_\mathbf{r}\rho \right|^2 + \frac{1}{2}T\kappa_\eta \left| \nabla_\mathbf{r}\eta \right|^2\right] \text{d}r, \\
& \gamma_\text{gb}=\int_{-\infty}^\infty \left[ f(T,\rho=1,\{\eta_\alpha,\eta_\beta\}) + \frac{1}{2}T\kappa_\eta \left| \nabla_\mathbf{r}\eta_\alpha \right|^2 + \frac{1}{2}T\kappa_\eta \left| \nabla_\mathbf{r}\eta_\beta \right|^2\right] \text{d}r
\end{split}
\label{eq29}
\end{equation}
When the equilibrium is reached, an application of the Euler-Lagrange equation to Eq. \ref{eq29} leads to
\begin{equation}
\begin{split}
& \gamma_\text{sf}=2\sqrt{\frac{T(\kappa_\eta+\kappa_\rho)}{2}} \int_0^1 \sqrt{f_\text{ht}\rho + (\underline{C}+7\underline{D})(1-\rho)^2\rho^2} \text{d}\rho, \\
& \gamma_\text{gb}=2\sqrt{T\kappa_\eta} \int_0^1 \sqrt{f_\text{ht} + 12\underline{D}(1-\rho)^2\rho^2} \text{d}\eta ,
\end{split}
\label{eq30}
\end{equation}
with $f_\text{ht}(T)=-c_\text{r}T\text{ln}(T/T_0)+c_\text{r}(T-T_0)$, $\underline{C}=\underline{C}_\text{pt}-\underline{C}_\text{cf}(T-T_0)$, and $\underline{D}=\underline{D}_\text{pt}-\underline{D}_\text{cf}(T-T_0)$.
Detailed derivations are shown in \ref{appendixa}. Eqs. (\ref{eq29}) and (\ref{eq30}) also present the temperature dependencies of $\gamma_\text{sf}$ and $\gamma_\text{gb}$. Since $\kappa_\rho$ and $\kappa_\eta$ and the relative specific heat $c_\text{r}$ are only material-dependent, such temperature dependencies are majorly inherited from the temperature-dependent free energy density $f(T,\rho,\{\eta_\alpha\})$. As for the physical picture conveyed from the formulation of $f(T,\rho,\{\eta_\alpha\})$ in Eq. (\ref{eq16}), there are contributions of both heat and the configuration to the variation of the free energy profile when temperature changes. In these model parameters, dimensionless parameters $\underline{A}$ and $\underline{B}$ basically show the proportion of the heat contribution, which is interpolated by different order parameters. $\underline{C}_\text{pt}$  and $\underline{D}_\text{pt}$ (with the dimension of the energy) determine $\gamma_\text{sf}$ and $\gamma_\text{gb}$. $\underline{C}_\text{cf}$ and $\underline{D}_\text{cf}$ (with the dimension of the entropy) determine the temperature dependency of $c_\text{r}$.

On the other hand, Eq. (\ref{eq30}) takes a degenerated form at $T_0$, where both $\gamma_\text{sf}^{T_0}$ and $\gamma_\text{gb}^{T_0}$ are only related to the gradient model parameters ($\kappa_\rho$ and $\kappa_\eta$) and the potential model parameters ($\underline{C}_\text{pt}$ and $\underline{D}_\text{pt}$), i.e.
\begin{equation}
\begin{split}
& \gamma_\text{sf}^{T_0}=\frac{1}{3\sqrt{2}} \sqrt{T_0(\kappa_\eta+\kappa_\rho)(\underline{C}_\text{pt}+7\underline{D}_\text{pt})}, \\
& \gamma_\text{gb}^{T_0}= \frac{2}{\sqrt{3}}\sqrt{T_0 \kappa_\eta \underline{D}_\text{pt}}.
\end{split}
\label{eq31}
\end{equation}
Considering the constraint for the model parameters in Eq. (\ref{eqA7}), we can also find that $\gamma_\text{sf}^{T_0}=\gamma_\text{gb}^{T_0}$ when $\kappa_\rho=\kappa_\eta$ and a dihedral angle of 120$^\circ$ occurs at $T_0$. Then, $\kappa_\rho > \kappa_\eta$ and $\kappa_\rho < \kappa_\eta$ characterize two kinds of model materials. The former has $\gamma_\text{sf}^{T_0}$ higher than $\gamma_\text{gb}^{T_0}$ and the dihedral angle over 120$^\circ$, while the later has $\gamma_\text{gb}^{T_0}$ higher  than $\gamma_\text{sf}^{T_0}$ and the dihedral angle below 120$^\circ$.

\begin{figure}[!t]
\centering
\includegraphics[width=17cm]{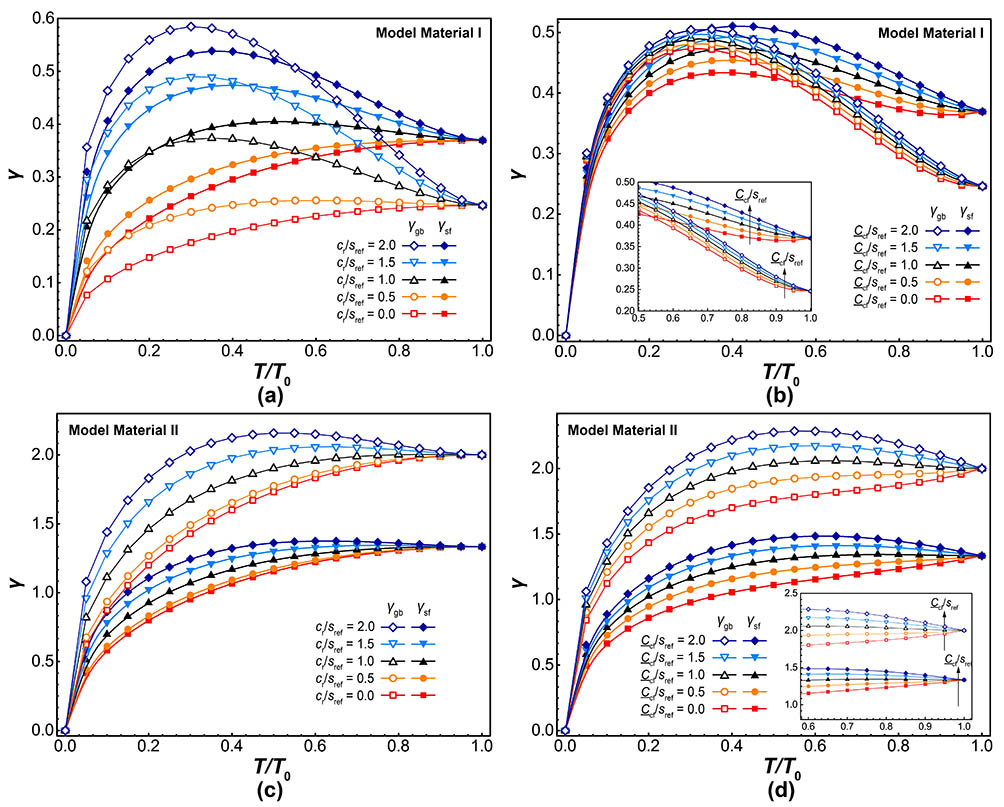}
\caption{ Temperature dependency of the surface and grain-boundary energies with different set of model parameters: (a) $\kappa_\rho =2 \kappa_\eta$, $\underline{C}_\text{cf}/s_\text{ref}=1$; (b) $\kappa_\rho =2 \kappa_\eta$, $c_\text{r}/s_\text{ref}=1$; (c) $3\kappa_\rho =\kappa_\eta$, $\underline{C}_\text{cf}/s_\text{ref}=1$; (d) $3\kappa_\rho =\kappa_\eta$, $c_\text{r}/s_\text{ref}=1$. The reference quantity $s_\text{ref}=\underline{C}_\text{pt}/T_0$}
\label{fig4}
\end{figure}

Fig. \ref{fig4} presents $\gamma_\text{sf}(T)$ and $\gamma_\text{gb}(T)$ as functions of the heat contribution ($c_\text{r}$) and the configuration contribution ($\underline{C}_\text{cf}$ and $\underline{D}_\text{cf}$) in the above-defined two model materials. To show the relative size of the $c_\text{r}$ and $\underline{C}_\text{cf}$, a reference quantity $s_\text{ref}$ was set as $\underline{C}_\text{pt}/T_0$ with the dimension of the entropy. In the first model material I (Fig. \ref{fig4}a and b), $\kappa_\rho =2 \kappa_\eta$ leads to $11\underline{D}_\text{pt}=\underline{C}_\text{pt}$ and $11\underline{D}_\text{cf}=\underline{C}_\text{cf}$ according to Eq. (\ref{eqA7}). Eventually, Eq. (\ref{eq31}) results in $\gamma_\text{sf}^{T_0}=1.5\gamma_\text{gb}^{T_0}$. In the model material II (Fig. \ref{fig4}c and d), $3\kappa_\rho =\kappa_\eta$ leads to $\underline{D}_\text{pt}=\underline{C}_\text{pt}$, $\underline{D}_\text{cf}=\underline{C}_\text{cf}$ , and finally $1.5\gamma_\text{sf}^{T_0}=\gamma_\text{gb}^{T_0}$. All the parameters are normalized according to Table \ref{tab1}. Fig. \ref{fig4} shows that both $\gamma_\text{sf}(T)$ and $\gamma_\text{gb}(T)$ start from zero at $T=0$, and end up with a certain positive value $\gamma_\text{sf}^{T_0}$ and $\gamma_\text{gb}^{T_0}$ at $T_0$. When $c_\text{r}=0$, both $\gamma_\text{sf}(T)$ and $\gamma_\text{gb}(T)$ present monotonic increasing trends from zero to $\gamma_\text{sf}^{T_0}(T)$ and $\gamma_\text{gb}^{T_0}(T)$, respectively. With the increase of $c_\text{r}$, a maximum emerges in the middle, making $\gamma_\text{sf}(T)$ and $\gamma_\text{gb}(T)$ increase in the low-temperature range ($0<T<0.5T_0$) then decrease in the high-temperature range ($0.5T_0<T<T_0$). As for increasing $\underline{C}_\text{cf}$, trends of two model materials are different. In the model material I with fixed $c_\text{r}/s_\text{ref}=1$, increasing $\underline{C}_\text{cf}$ leads to quickly decrease of $\gamma_\text{sf}(T)$ and $\gamma_\text{gb}(T)$ when temperature increases in the high-temperature range. In the model material II with fixed $c_\text{r}/s_\text{ref}=1$, however, increasing $\underline{C}_\text{cf}$ makes $\gamma_\text{sf}(T)$ and $\gamma_\text{gb}(T)$ decreases with the increasing temperature in the high-temperature range. Since $c_\text{r}$ can be directly obtained from the specific heat of the substance and the atmosphere, it is flexible to use the parameter $\underline{C}_\text{cf}$ to modify the temperature dependency of the $\gamma_\text{sf}(T)$ and $\gamma_\text{gb}(T)$. Another important detail is that $\gamma_\text{sf}(T)$ and $\gamma_\text{gb}(T)$ calculated by Eq. (\ref{eq30}) show an approximately linear trend in the high temperature range (over $0.6T_0$), agreeing well with the previous experimental results \citep{67tsoga1996surface,68zouvelou2007interfacial,69zouvelou2008surface}.
 
\subsection{Benchmark test of the dihedral angle}\label{bench}
In order to validate the model and its numerical implementation for capturing the grain coalescence during the sintering, benchmark simulation is firstly carried out. The measurement of the dihedral angle, which reflects the ratio of the surface energy to the grain-boundary energy at a certain temperature, is used as the benchmark problem \citep{1kang2004sintering,42ahmed2013phase,61moelans2008quantitative,65warren2003extending}. According to the Young’s law, the dihedral angel formed by two particles/grains at equilibrium reads as
\begin{equation}
\Phi=2\text{arccos}\frac{\gamma_\text{gb}}{2\gamma_\text{sf}}.
\label{eq28}
\end{equation}
$\gamma_\text{sf}$ and $\gamma_\text{gb}$ are the surface and grain-boundary energy, respectively. Their temperature dependencies have been revealed in Subsection \ref{gbenergy}. Thus, it is straightforward to analytically calculate the temperature-dependent dihedral angle $\Phi$ by using Eq. (\ref{eq28}).

\begin{figure}[!t]
\centering
\includegraphics[width=14cm]{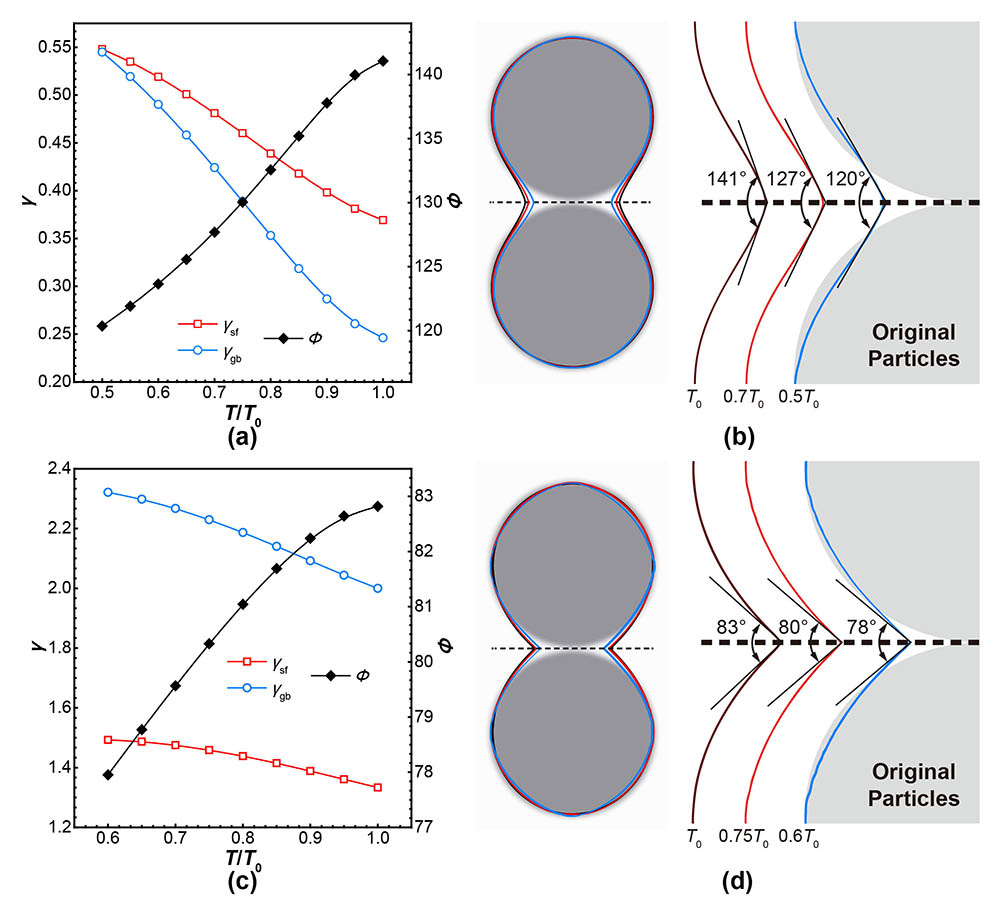}
\caption{Temperature-dependent dihedral angle: (a) analytical and (b) simulated dihedral angle of the model material I; (c) analytical and (d) simulated dihedral angle of the  material II.}
\label{fig5}
\end{figure}

We consider the grain coalescence from two identical particles at various temperatures. Fig. \ref{fig5} shows the dihedral angle results from the analytic solution in Eq. (\ref{eq28}) and the FEM simulations. The model with a square domain $100\lambda \times 100\lambda$ was constructed ($\lambda$ is the reference length). Two identical particles with a radius of $20\lambda$ were placed in the center. One particle had the orientation $\eta_1$  and the other $\eta_2$. Periodic boundary condition was set for the order parameters $\rho$ and $\{\eta_\alpha\}$ on both directions while temperature $T$ was fixed on every node to make a strictly isothermal condition. Two kinds of materials mentioned in Subsection \ref{gbenergy} (Fig. \ref{fig4}) were utilized, i.e. model material I with $\kappa_\rho=2\kappa_\eta$ and model material II with $2\kappa_\rho=\kappa_\eta$. For both model materials, we assigned $c_\text{r}/s_\text{ref}=1.5$ and $C_\text{cf}/s_\text{ref}=1.5$. The normalized diffusive mobility is set as $\widetilde{M}_{ij}=10\widetilde{L}$. After running the simulation with the total time $t^*=200\tau$ to reach the equilibrium, we captured the morphology of the sintered particles and measured the dihedral angle, as shown in Fig. \ref{fig5}b and d. In contrast, Fig. \ref{fig5}a and c present the dihedral angle calculated from Eq. (\ref{eq28}). It can be seen from Fig. 5 that the FEM simulation results agree well with the analytic solutions, demonstrating the validation of the FEM implementation.

\subsection{Grain coalescence from two identical particles with temperature gradient}\label{twoinden}
\begin{figure}[!t]
	\centering
	\includegraphics[width=16cm]{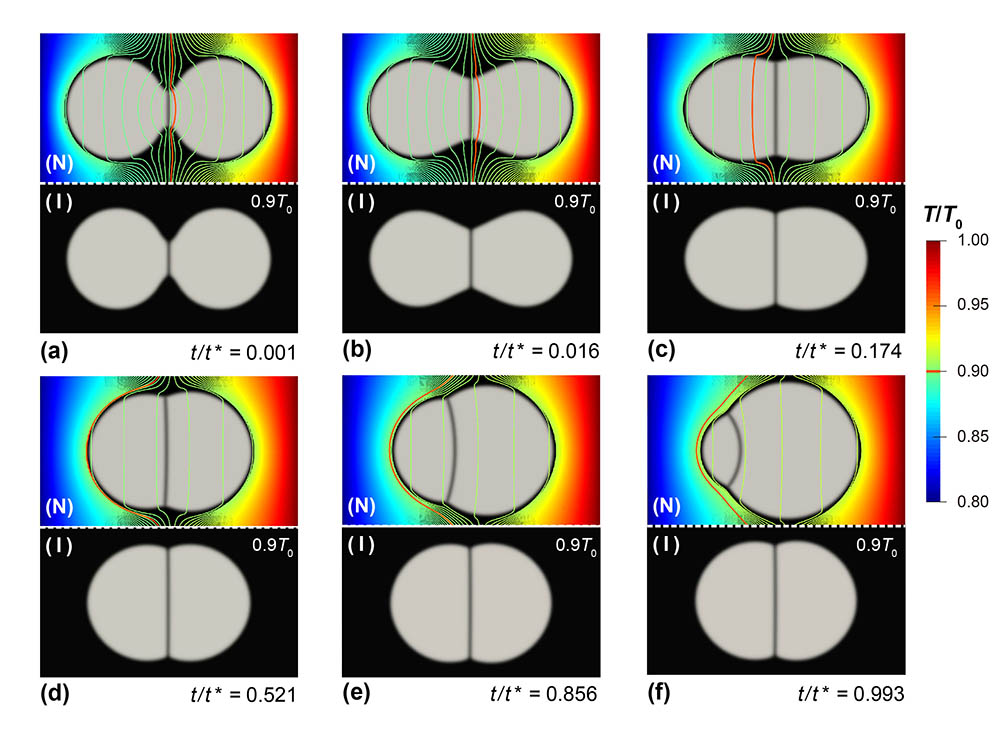}
	\caption{Comparison of the snapshots between the isothermal grain coalescence (I) and the non-isothermal grain coalescence (N) model with temperature gradient parallel to the common axis of two particles.
Detailed temporal evolution can be found in Movie S1 in the Supplementary Information.}
	\label{fig6}
\end{figure}

According to the discussion in the Subsection \ref{free}, grains with different orientations should be thermodynamically equally stable on the isothermal condition, since a symmetric "multi-well" formulation is employed for the free energy density. However, when there is the temperature gradient inside the sintering system, such multiple wells will be tilted, resulting in different thermodynamic stability among grains. For a validation, comparison between the grain coalescence from two identical particles with and without the temperature gradient was carried out. The model material I ($\kappa_\rho=2\kappa_\eta$, $c_\text{r}/s_\text{ref}=1.5$, and $\underline{C}_\text{cf}/s_\text{ref}=1.7$) was utilized for two particles with a radius of $20\lambda$. Normalized surface diffusive mobility was set as $\widetilde{M}_{ij}=10\widetilde{L}$. To distinguish the heat transfer in the bulk and the atmosphere/pores, normalized thermal conductivity of the bulk was set as $\widetilde{k}_\text{bk}=10^6\widetilde{L}$ and that of the atmosphere/pores as  $\widetilde{k}_\text{at}=0.01\widetilde{k}_\text{bk}$. We set the boundary conditions as $T_\text{left}=0.8T_0$ and $T_\text{right}=T_0$ to induce a temperature gradient parallel to the common axis of two particles for the non-isothermal case, as shown in Fig. \ref{fig6}. On the contrary,  the temperature $T$ was fixed at $0.9T_0$ on every node for the isothermal case. The simulation was run for a total time of $t^*=10^4 \tau$. Results in Fig. \ref{fig6} show the shrinkage and neck growth between two identical particles before $t/t^*=0.521$. For the isothermal case, the morphology of the sintered particles always remains symmetric. And the grain boundary is always a straight line and stays in the center. This symmetric morphology has also been predicted by the previous work \citep{61moelans2008quantitative,70zhang1995sintering}. In contrast, for the non-isothermal case, this symmetric morphology is broken and the grain boundary is curved after $t/t^*=0.521$. Meanwhile, the grain at the hot side gradually grows bigger than that at the cold side. The hotter grain shows the tendency to completely merge the colder one.

\begin{figure}[!t]
	\centering
	\includegraphics[width=16cm]{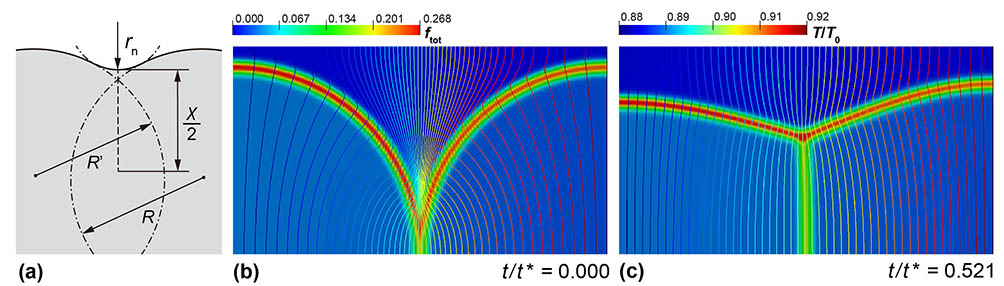}
	\caption{(a) Approximated geometries around the neck: the particle/grain with radius $R^\prime$ located at the cold side and $R$ located at the hot side. The isotherms (temperature contours) and the total free energy density $f^\text{tot}$ at (b) $t/t^*=0$ and (c) $t/t^*=0.521$.}
	\label{fig7}
\end{figure}

The different morphology in isothermal and non-isothermal grain coalescence can be understood as follows. According to Young-Laplace equation, the local chemical potential on the surface and interface can be formulated as \citep{1kang2004sintering,71somiya2013handbook}
\begin{equation}
\hat{\mu}=\gamma^\text{H} H V_\text{m},
\label{eq32}
\end{equation}
where $\gamma^\text{H}$ is the on-site surface and interface energy, $H$ the main curvature (notice that here we only discuss 2D case), and $V_\text{m}$ the molar volume. The curvature radius around the neck is $r_\text{n}$. The radii of two particles $R$ and $R^\prime$ are assumed to keep identical during the shrinkage. Based on the approximated geometries in Fig. \ref{fig7}a, then the chemical potential differences for the neck growth $\Delta_\text{ng}\hat{\mu}$ and for the grain growth $\Delta_\text{gg}\hat{\mu}$ can be given as \citep{1kang2004sintering,43ahmed2014phase,71somiya2013handbook}
\begin{equation}
\Delta_\text{ng}\hat{\mu}=V_\text{m}\left(\frac{\gamma_\text{sf}^r}{r_n} - \frac{\gamma_\text{sf}^R}{R}\right) ~ \text{or} ~ V_\text{m}\left(\frac{\gamma_\text{sf}^r}{r_n} - \frac{\gamma_\text{sf}^{R^\prime}}{R^\prime}\right),
\label{eq33}
\end{equation}
\begin{equation}
\Delta_\text{gg}\hat{\mu}=V_\text{m}\left(\frac{\gamma_\text{gb}^{R^\prime}}{R^\prime} - \frac{\gamma_\text{gb}^R}{R}
\label{eq34}\right)
\end{equation}
For the isothermal grain coalescence, $\gamma_\text{sf}^r=\gamma_\text{sf}^R$ (or $\gamma_\text{sf}^{R^\prime}$) and $\gamma_\text{gb}^{R^\prime}=\gamma_\text{gb}^R$ should always hold. Then only the curvatures determine the chemical potentials. Since initially $r_\text{n} \ll R$, $\Delta_\text{ng}\hat{\mu}$ is always larger than $\Delta_\text{gg}\hat{\mu}$, making the neck growth dominant firstly and then slow down when $r_\text{n} \rightarrow R$ (or $R^\prime$). On the other hand, since $R$ and $R^\prime$ keep identical during shrinkage, $\Delta_\text{gg}\hat{\mu}=0$. It means that there is vanishing driving force for the grain growth, and thus a straight grain boundary would be ideally formed. 

For the non-isothermal grain coalescence, however, the on-site surface and grain-boundary energies could be non-identical due to the existence of the temperature gradient. Fig. \ref{fig7}b and c show the region around the neck with isotherms and the total free energy density $f_\text{tot}$ which is calculated as
\begin{equation}
f_\text{tot}=f+\frac{1}{2}T\kappa_\rho \left|\nabla_\mathbf{r}\rho \right|^2 + \frac{1}{2}T\kappa_\eta \left|\nabla_\mathbf{r}\eta \right|^2.
\label{eq35}
\end{equation}
Then the on-site surface and grain-boundary energies can be estimated by using $f_\text{tot}$ according to Eq. (\ref{eq29}). In this case, we find the isotherms are perpendicular to the surface while almost parallel to the grain boundary, i.e. $\gamma_\text{sf}^r \neq \gamma_\text{sf}^R$ but $\gamma_\text{gb}^{R^\prime} \approx \gamma_\text{gb}^R$. At the beginning, isotherms are rather concentrated around the neck, indicating a high temperature gradient. But due to $r_\text{n} \ll R$, the neck growth is still dominant at the initial stage. Around $t/t^*=0.521$ when $r \rightarrow R$ (or $R^\prime$) due to the temperature gradient on the surface, we can approximately find $\gamma_\text{sf}^{R^\prime}<\gamma_\text{sf}^r<\gamma_\text{sf}^R$. Therefore, positive $\Delta \hat{\mu}$ can be found through the route "colder grain $\rightarrow$ neck $\rightarrow$ hotter grain", which drives the mass transportation (e.g. surface diffusion) and results in gradual growth of the grain at the hot side. Once the grains become non-identical, i.e. $R\neq R^\prime$, non-zero $\Delta_\text{gg}\hat{\mu}$ emerges and grain growth starts.

\subsection{Grain coalescence from two non-identical particles with multiple influence factor}\label{twononiden}
\begin{figure}[!t]
\centering
\includegraphics[width=16cm]{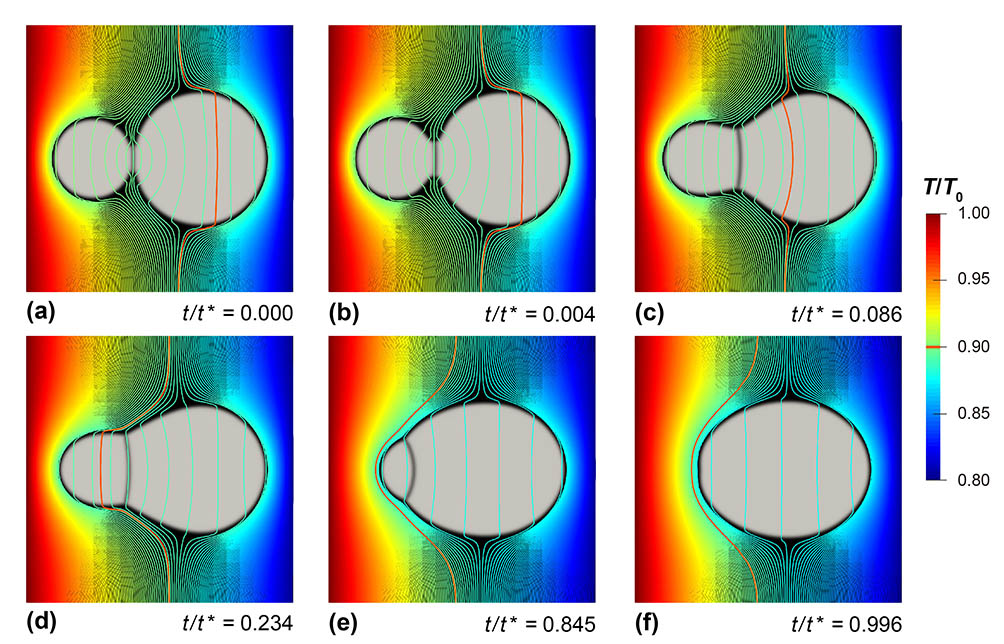}
\caption{Snapshots of the non-isothermal grain coalescence from two non-identical particles with temperature gradient parallel to the common axis of two particles. Isotherms are also illustrated, with $0.9T_0$ isotherm highlighted.
Detailed temporal evolution can be found in Movie S2 in the Supplementary Information.}
\label{fig8}
\end{figure}

\begin{figure}[!t]
	\centering
	\includegraphics[width=17cm]{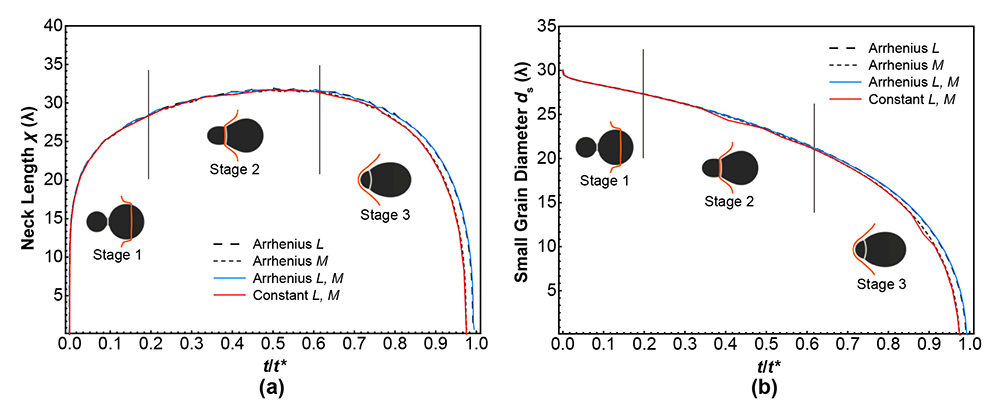}
	\caption{Effects of diffusive mobility (M) and grain-boundary mobility (L) on (a) neck length and (b) small-grain diameter.}
	\label{fig9}
\end{figure}

In Fig. \ref{fig8} we present the snapshots of non-isothermal grain coalescence from two non-identical particles, where typical neck growth and shrinkage occur in the small grain. The initial radius of the large and small grain is $R_\text{s}=20\lambda$ and $R_l=20 \lambda$, respectively. Other parameters and boundary conditions were set the same as in Subsection \ref{twoinden}. The effective neck length ($\chi$) and effective diameter of the small grain ($d_\text{s}$) are calculated by (suppose $\eta_\alpha$ corresponds to the small grain)
\begin{equation}
\chi=\int_\Omega\frac{2\eta_\alpha\eta_\beta}{\widetilde{\lambda}_\text{gb}} \text{d}\Omega \quad \text{and} \quad d_\text{s}=\frac{2}{\pi}\sqrt{\int_\Omega \eta_\alpha \text{d}\Omega},
\label{eq36}
\end{equation}
where $\widetilde{\lambda}_\text{gb}$ is the normalized grain-boundary width, i.e. $\widetilde{\lambda}_\text{gb}=\sqrt{4\widetilde{\kappa}_\eta /3 \widetilde{\underline{D}}_\text{pt}}$. Based on the temporal evolution of $\chi$ and $d_\text{s}$ in Fig. \ref{fig9}, we can divide the non-isothermal grain coalescence process of two non-identical particles into three stages. We consider $\Delta_\text{ng}\hat{\mu}$ for neck growth and  $\Delta_\text{gg}\hat{\mu}$ for the grain growth in Eqs. (\ref{eq33}) and (\ref{eq34}). In the first stage when the curvature radius of the neck ($r_\text{n}$) is still small, the neck growth predominates, which is featured by the rapid increase of $\chi$ but rather slow decrease of $d_\text{s}$. In the second stage when $r_\text{n}$ and $\chi$ approach the magnitude of the small $d_\text{s}$, both $\chi$ and $d_\text{s}$ change slowly and a curved grain boundary forms. In the third stage, the small grain quickly shrinks due to the migration of grain boundary (the grain growth dominates). This then results in the decease of  $\chi$ and $d_\text{s}$, i.e. the reduction of the surface and grain boundary and thus the total free energy. Such three-stage pictures  here are coherent with the analytical description in \citep{19lange1989thermodynamics,72kellett1989thermodynamics}.

Many factors would influence the grain coalescence. Among them, the temperature-dependent diffusivity and grain-boundary mobility vary drastically with respect to the temperature change and play a critical role. The effective diffusivity $D_p^\text{eff}$ of the process $p$ ($p=$ bk, vp, sf, gb as shown in Eq. (\ref{eq19})) and grain-boundary mobility $G^\text{eff}_\text{gb}$ (Eq. (\ref{eq21})) at certain temperature $T$ usually obey the Arrhenius equation, i.e.
\begin{equation}
D_p^\text{eff}(T)=D_p^\text{eff}(T_0) \text{exp}\left[-\frac{\mathscr{E}_p^\text{D}}{\mathfrak{R}}\left(\frac{1}{T} -\frac{1}{T_0}\right)\right] \quad \text{and} \quad
G_\text{gb}^\text{eff}(T)=G_\text{gb}^\text{eff}(T_0) \text{exp}\left[-\frac{\mathscr{E}_{\text{gb}}^\text{G}}{\mathfrak{R}}\left(\frac{1}{T} -\frac{1}{T_0}\right)\right]
\label{eq37},
\end{equation} 
where $D_p^\text{eff}(T_0)$ and $G_\text{gb}^\text{eff}(T_0)$ are the total effective diffusivity and grain-boundary mobility measured at $T_0$ with corresponding activation energy $\mathscr{E}_p^\text{D}$ and $\mathscr{E}_{\text{gb}}^\text{G}$ in J/mol, respectively. $\mathfrak{R}$ is the ideal gas constant. In this model, we use the modified mobilities $M_{ij}(T)=D_{ij}(T) / \left[2\left(\underline{C}_\text{pt}+\underline{D}_\text{pt} \right) \right]$ and $L(T)=G_\text{gb}^\text{eff}(T)\gamma^{T}_\text{gb}/(T\kappa_\eta)$, in which the total diffusivity $D_{ij}$ follows Eq. (\ref{eq18}). Then we simply formulate both $M_{ij}$ and $L$ in the Arrhenius form, which after normalization reads
\begin{equation}
\widetilde{M}_{ij}(\widetilde{T})=\widetilde{M}_{ij}(T_0) \text{exp}\left[-\frac{\mathscr{E}_{ij}^\text{D}}{\mathfrak{R}T_0} \left(
\frac{1}{\widetilde{T}} -1 \right)\right] \quad \text{and} \quad
\widetilde{L}(\widetilde{T})=\widetilde{L}(T_0) \frac{\widetilde{\gamma_{\text{gb}}}}{{\widetilde{T}}}\text{exp}\left[-\frac{\mathscr{E}_{\text{gb}}^\text{G}}{\mathfrak{R}T_0} \left(
\frac{1}{\widetilde{T}} -1 \right)\right],
\label{eq38}
\end{equation}
where $\widetilde{\gamma_{\text{gb}}} = \gamma^{T}_\text{gb} / \gamma^{T_0}_\text{gb} $ and $\widetilde{T}=T/T_0 $. In the case we get $\widetilde{M}_{ij}(T_0)=10\widetilde{L}(T_0)$ and $\mathscr{E}_{\text{gb}}^\text{G}=\mathscr{E}^\text{D}_{ij}=\mathfrak{R}T_0$, then set up the simulations with (i) only $\widetilde{L}(\widetilde{T})$ in Arrhenius form, $\widetilde{M}_{ij}(\widetilde{T})=\widetilde{M}_{ij}(T_0)$; (ii) only $\widetilde{M}_{ij}(\widetilde{T})$ in Arrhenius form, $\widetilde{L}(\widetilde{T})=\widetilde{L}(T_0)$; (iii) both $\widetilde{L}(\widetilde{T})$ and $\widetilde{M}_{ij}(\widetilde{T})$ in Arrhenius form; and (iv) both $\widetilde{L}(\widetilde{T})$ and $\widetilde{M}_{ij}(\widetilde{T})$ as constant, $\widetilde{L}(\widetilde{T})=\widetilde{L}(T_0)$, $\widetilde{M}_{ij}(\widetilde{T})=\widetilde{M}_{ij}(T_0)$. Results in Fig. \ref{fig9}a and b indicate that there is very little difference among simulations with different $\widetilde{L}(\widetilde{T})$ and $\widetilde{M}_{ij}(\widetilde{T})$ in the first two stages. In the third stage, however, for simulations with $\widetilde{L}(\widetilde{T})$ in Arrhenius form, it takes relatively longer time to reach the finial equilibrium. Considering the relative position of the $0.9T_0$ isotherm (highlighted in Fig. \ref{fig8}) in different stages, we find that the neck and its surrounding region are completely at the hot side of the $0.9T_0$ isotherm in the first stage, i.e. $0.89\widetilde{L}(T_0)<\widetilde{L}(\widetilde{T})<\widetilde{L}(T_0)$ and $0.89\widetilde{M}_{ij}(T_0)<\widetilde{M}_{ij}(\widetilde{T})<\widetilde{M}_{ij}(T_0)$. Both grains and the grain boundary are completely at the cold side of the $0.9T_0$ isotherm in the third stage, i.e. $\widetilde{L}(\widetilde{T})<0.89\widetilde{L}(T_0)$ and $\widetilde{M}_{ij}(\widetilde{T})<0.89\widetilde{M}_{ij}(T_0)$. These imply that smaller $\widetilde{L}(\widetilde{T})$ and $\widetilde{M}_{ij}(\widetilde{T})$ obviously delay the process of grain growth in the third stage, but have limited influence on the neck growth in the first and second stages.

\begin{figure}[!t]
	\centering
	\includegraphics[width=16cm]{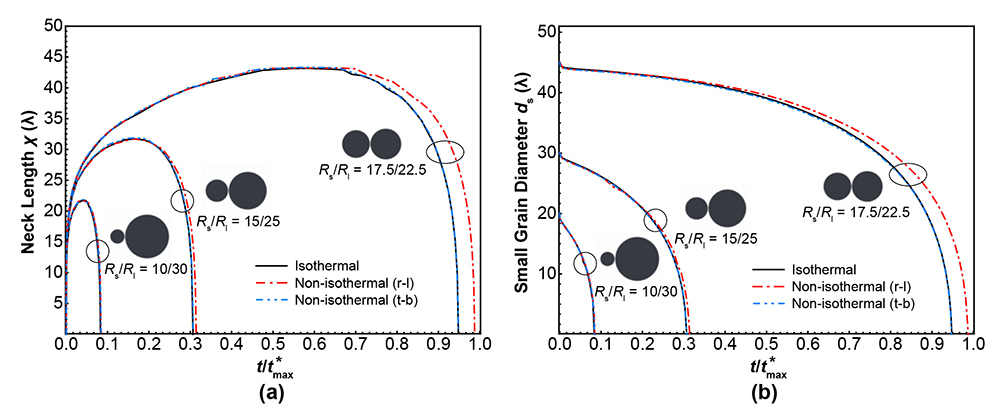}
	\caption{Effects of initial particle size difference on (a) neck length and (b) small-grain diameter.}
	\label{fig10}
\end{figure}

The size difference of the initial particles, i.e. the ratio of size between small and large particles, can also influence the grain coalescence. Fig. \ref{fig10}a and b present the neck growth and the small grain shrinkage during the grain coalescence from two non-identical particles with different $R_\text{s}/R_1$. Basically, small $R_\text{s}/R_1$ will accelerate the grain coalescence. When high $R_\text{s}/R_1$, especially when $R_\text{s}/R_1 \rightarrow 1$, the system will be more easily affected by the temperature gradient parallel to the common axis of two particles. The gradient perpendicular to the common axis barely influences the system. This can be explained as follows. The curvature radius of the neck ($r_\text{n}$) approaches the magnitude of  $R_\text{s}$ at the end of the neck growth. If $R_\text{s}/R_1 \rightarrow 1$, the effect of on-site surface energy difference (caused by the temperature gradient parallel to the common axis) on $\Delta \hat{\mu}$  would  dominate, while the effect of the curvature would be less obvious. Hence, the mass transportation, e.g. surface diffusion, would be more likely to occur through the route "colder grain $\rightarrow$ neck $\rightarrow$ hotter grain".

\subsection{Multi-particle coalescence informed by experimental data}\label{multipart}
In this section we present a more practical example, i.e. multi-particle system, based on the experimental data on ceria (CeO$_2$), which is one of the most promising electrolyte ceramics for solid oxide fuel cell system \citep{473inaba1996ceria,474kleinlogel2000sintering}. Ceria is a fluorite material with relatively high melting point (around 3000 K) \citep{73li2004low} and is available to modify the sintering temperature in a wide range through doping or nanoparticle sintering \citep{69zouvelou2008surface,474kleinlogel2000sintering,73li2004low,476kleinlogel2001sintering}. Porosity, as well as its morphology during the sintering, is also of general interests \citep{474kleinlogel2000sintering,477wang2009porous}. There are several previous reports on the temperature-dependent properties of ceria, in particular, the surface and grain-boundary energy \citep{69zouvelou2008surface}, surface and bulk ionic diffusivities \citep{69zouvelou2008surface,478ruiz1998oxygen}, and grain-boundary mobility \citep{80chen1996grain}. The non-isothermal grain coalescence simulation of multiple particles will be informed by these experimental results. Table \ref{tab2} gives the experimental parameters of ceria with argon atmosphere in the temperature range of 1473--1773 K.

\begin{figure}[!t]
\centering
\includegraphics[width=16cm]{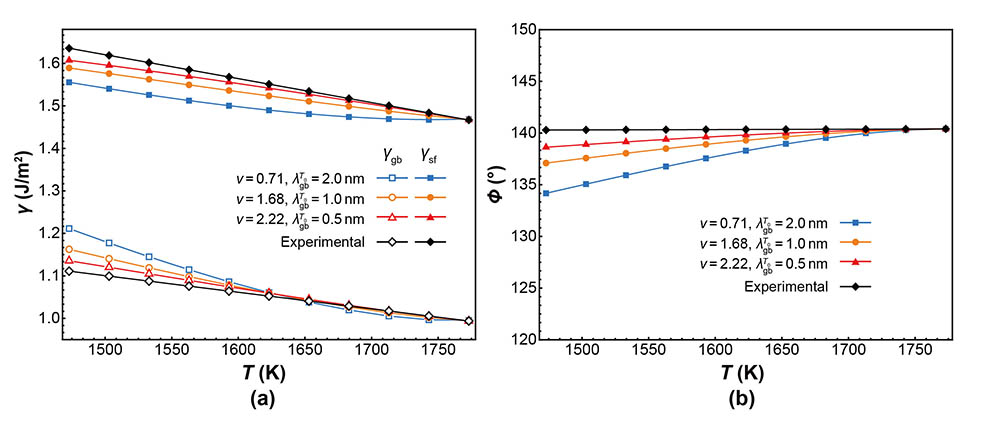}
\caption{Comparison between the model value and the experimental value of (a) surface and grain-boundary energies and (b) dihedral angle with various $\lambda_\text{gb}^{T_0}$ and $\nu$. The parameter $\nu$ is defined as $\nu=T_0\underline{D}_\text{cf}/\underline{D}_\text{pt}=T_0\underline{C}_\text{cf}/\underline{C}_\text{pt}$ according to Eq. (\ref{eqA7b}).}
\label{fig11}
\end{figure}

\renewcommand{\arraystretch}{1.5}
\begin{table}[!t] \small
\centering 
\caption{Material parameters used in simulations.}
\begin{tabular}{cccccc}
\hline
Properties & Expressions ($T$ in K) & Units &  References \\ \hline
$\gamma_\text{sf}$ & $2.465-0.563\times 10^{-3}T$ & J/m$^2$ & \cite{69zouvelou2008surface} \\
$\gamma_\text{gb}$ & $1.68-0.391\times 10^{-3}T$ & J/m$^2$ & \cite{69zouvelou2008surface} \\
$D_\text{sf}^\text{eff}$ & $3.82\times 10^{-4}\text{exp}\left(-0.308\times 10^5/\mathfrak{R}T \right)$ & m$^2$/s & \cite{69zouvelou2008surface} \\
$G_\text{gb}^\text{eff}$ & $8.7\times 10^4\text{exp}\left(-5.89\times 10^5/\mathfrak{R}T \right)$ & m$^4$/(Js) & \cite{80chen1996grain} \\
$k_\text{ceria}^\text{eff}$ & $1/\left(6.776\times 10^{-2} + 2.793\times 10^{-4}T \right)$ & J/(smK) & \cite{81nelson2014evaluation} \\
$c_\text{ceria}$ & $\sim 0.5\times 10^3$ & J/(kgK) & \cite{81nelson2014evaluation} \\
    \hline
\end{tabular}
\label{tab2}
\end{table}

\begin{figure}[!t]
\centering
\includegraphics[width=16cm]{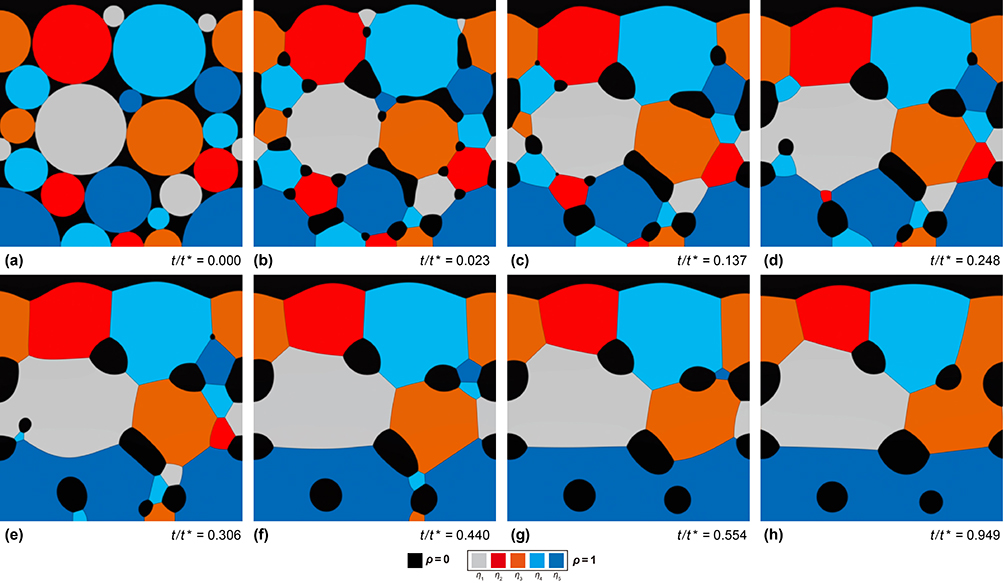}
\caption{Grain coalescence during the non-isothermal grain coalescence process.
Detailed temporal evolution can be found in Movie S3 in the Supplementary Information.}
\label{fig12}
\end{figure}

\begin{figure}[!t]
\centering
\includegraphics[width=16cm]{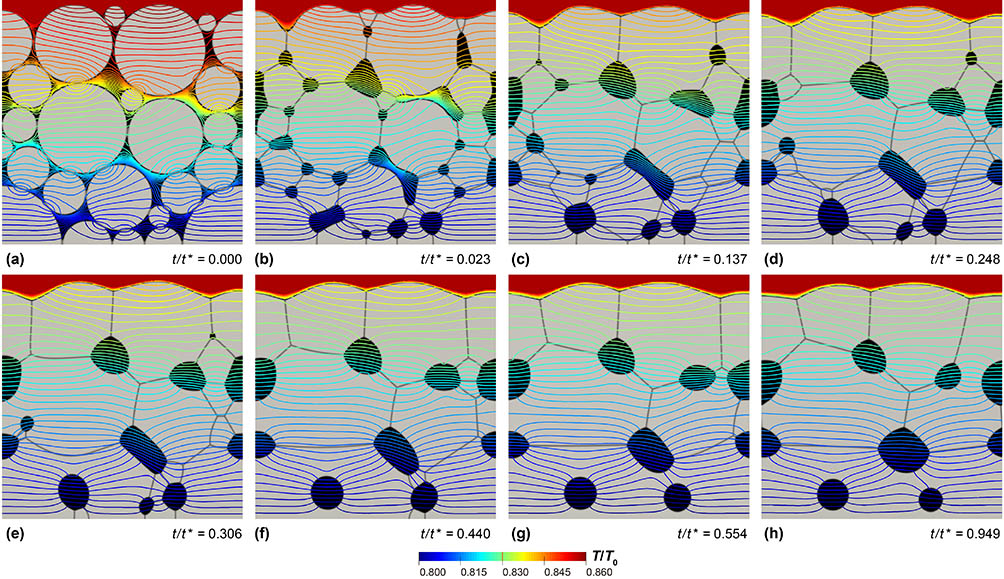}
\caption{Temporal evolution of isotherms during the non-isothermal grain coalescence process.}
\label{fig13}
\end{figure}

\begin{figure}[!t]
\centering
\includegraphics[width=10cm]{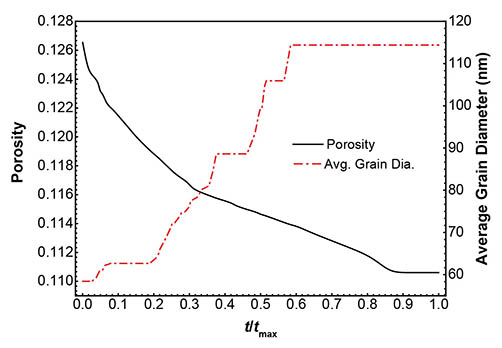}
\caption{Porosity and the average grain diameter as a function of time during the non-isothermal grain coalescence process.}
\label{fig14}
\end{figure}

There are eight parameters ($\underline{A}$, $\underline{B}$, $\underline{C}_\text{pt}$, $\underline{C}_\text{cf}$, $\underline{D}_\text{pt}$, $\underline{D}_\text{cf}$, $\kappa_\rho$, $\kappa_\eta$) in the model.
Due to the four relations in Eq. (\ref{eqA7b}), there are only four independent quantities left, i.e. $\underline{D}_\text{pt}$, $\underline{D}_\text{cf}$, $\kappa_\rho$, $\kappa_\eta$ (or $\underline{C}_\text{pt}$, $\underline{C}_\text{cf}$, $\kappa_\rho$, $\kappa_\eta$). These independent quantities can be obtained by using the direct or indirect methods which have been explicitly discussed in the \ref{appendixa}. Here we use the indirect method. The  grain-boundary thickness $\lambda_\text{gb}^{T_0}$ at the reference temperature $T_0$ was estimated, then the parameter $\nu=T_0\underline{D}_\text{cf}/\underline{D}_\text{pt}=T_0\underline{C}_\text{cf}/\underline{C}_\text{pt}$ was fitted non-linearly. Fig. \ref{fig11}a shows the fitted $\gamma_\text{sf}$ and $\gamma_\text{gb}$ by using different sets of $\gamma_\text{gb}^{T_0}$ ($T_0=1773$ K) and $\nu$, as well as the experimental $\gamma_\text{sf}$ and $\gamma_\text{gb}$. The corresponding dihedral angel is shown in Fig. \ref{fig11}b. It can be found that when   decreases from 2 nm to 0.5 nm, the fitted  $\gamma_\text{sf}$ and $\gamma_\text{gb}$ are much closer to the experimental values, and the deviation of the dihedral angle $\Phi$ at the cold end decreases. It is expectable that a better fitting shall be achieved if a smaller $\lambda_\text{gb}^{T_0}$ is used. In fact, the experimental investigation in \cite{476kleinlogel2001sintering} show a grain-boundary thickness less than 0.5 nm at 1673 K. However, a smaller grain-boundary thickness would increase the calculation cost, since a finer mesh structure with more elements and thus more nodes are needed to resolve the grain boundary. Here, $\lambda_\text{gb}^{T_0}$ of 1 nm is used in the following simulation. In this case, deviations of 2.4$\%$, 4.5$\%$ and 3.5$\%$ on $\gamma_\text{sf}$, $\gamma_\text{gb}$, and $\Phi$ should occur, respectively.

The simulated domain is with a size of $400\times 400$ nm$^2$. The particle radius ranges from 15--100 nm. A temperature difference of $0.06T_0$ (about 100 K) between the top (argon atmosphere) and the bottom of the domain is set as boundary condition to simulate the non-isothermal grain coalescence. The total simulation time $t^*=5\times10^4 \tau$. The  non-isothermal grain coalescence during the powder-bed sintering process is shown in Fig. \ref{fig12} with pores and orientation fields.  The temporal evolution of porosity and the average grain size are presented in Fig. \ref{fig14}. These results provide details to characterize the different stages during the grain coalescence \citep{1kang2004sintering,2olevsky1998theory,3german1996sintering,4wong1979models}: in the first stages from Fig. \ref{fig12}a to c, we can clearly see the neck growth and shrinkage among grains, resulting in a bulk with multiple large and interconnected pores. In the second stages from Fig. \ref{fig12}d to f, the small grains continue shrinking to vanish due to the microstructure relaxation through the grain growth. Meanwhile, irregular pores with multiple surfaces reshape along with the grain growth, resulting in stabler pores with triple surfaces. In the third stages from Fig. \ref{fig12}g to h, some triple-surface pores vanish, and then the triple-grain junctions can be observed. Meanwhile, the spherical isolated pores emerge, making them more stable than the pores located on the grain boundaries. These close pores inhibit the further coarsening of the bulk. 

Due to the differences in thermal conductivity and heat capacity between bulk and atmosphere/pores, denser isotherms can be found in such regions, as shown in Fig. \ref{fig13}. If two pores locate very lose to each other, denser isotherms appear in the bulk, i.e. large local temperature gradient. Such local temperature gradient would cause the gradient of on-site surface and grain-boundary energies nearby. Therefore, the driving force of possible local mass transportations will be increased and eventually make those pores merge quickly. In contrast to the spherical and triple-boundary pores, the irregular pores are more likely to form locally  denser isotherms. This shall be also one of the reasons for their reshaping in the second stages.


\section{Conclusions}\label{conclu}
We have established a thermodynamically consistent non-isothermal phase-field model for the study of the non-isothermal grain coalescence process which is critically important in the unconventional sintering techniques such as the spark plasma sintering, field-assisted sintering, and selective laser sintering. We use conserved order parameters (i.e. the fractional density field representing the bulk and atmosphere/pore region) and the non-conserved ones (i.e. the orientation fields associated with crystallographic orientations) to describe the sintering system. The model derivation starts from the entropy functional, followed by the formulation of temperature dependent free energy density, kinetics for order parameters and the order-parameter-coupled heat transfer. Finite element implementation makes the model applicable to the practical non-isothermal grain coalescence from multiple particles. The main conclusions of this study are highlighted in the following:

1) The free energy density formulation in Eq. (\ref{eq16}) includes the internal energy (induced by the change of temperature and order parameters) and the order parameter related configurational entropy. It is designed to reflect the temperature-dependent equilibrium state. The temperature gradient during the non-isothermal grain coalescence is shown to break the symmetry of the free energy density distribution. The connection between the model parameters and the experimental measurement is also addressed.

2) In the non-isothermal grain coalescence, local temperature gradient induces the gradient of on-site surface and grain-boundary energies. Then positive driving force for the local mass transportation around the neck will occur along the path "colder grain $\rightarrow$ neck $\rightarrow$ hotter grain", leading to grain growth at the hotter side and unsymmetric morphology of two identical particles.

3) Based on the non-isothermal grain coalescence from two non-identical particles, the effects of particle size difference and temperature-dependent diffusive/grain-boundary mobilities on the temporal evolution of neck length and grain diameter are discussed. Also, a three-stage feature of the non-isothermal grain coalescence and the associated mechanism are identified.

4) Taking ceria (CeO$_2$) as a model material, we determine the model parameters by using the experimental data of CeO$_2$, and carry out phase-field simulations on the non-isothermal grain coalescence from multiple CeO$_2$ particles. The temporal grain coalescence and time-dependent average porosity and grain size are readily predicted.

The model presented in this study may provide a method for the simulation of unconventional sintering, in which the  non-equilibrium and high temperature gradient play a critical role. With appropriate modification by introducing laser-particle interactions, the model could be directly employed to simulate the grain coalescence in the unconventional sintering techniques. We leave these for future investigations.

\section*{Acknowledgments}
The support from the European Research Council (ERC) under the European Union´s Horizon 2020 research and innovation programme (grant agreement No 743116) and the Profile Area From Material to Product Innovation -- PMP (TU Darmstadt) is acknowledged. The authors also greatly appreciate their access to the Lichtenberg High Performance Computer of Technische Universit\"at Darmstadt.
L.-Q. Chen acknowledged the Humboldt Research Award.

\section*{Appendix}
\appendix
\setcounter{figure}{0}    
\renewcommand\thefigure{A\arabic{figure}}
\section{Explicit formulation of surface and grain-boundary energies at equilibrium and the determination of model parameters}\label{appendixa}
In this appendix we present the derivation of the surface and grain-boundary energies at equilibrium which is based on the previous works by \cite{61moelans2008quantitative}, \cite{474kleinlogel2000sintering}, and \cite{35fan1997effect}. Here we firstly show the dependency of the model parameters from Eq. (\ref{eqA1}) to (\ref{eqA8}), then derive the explicit formulation of the surface and grain-boundary energies from Eq. (\ref{eqA9}) to (\ref{eqA17}). Finally, direct and indirect methods of determining eight model parameters are proposed in the rest of this appendix.

\begin{figure}[!t]
	\centering
	\includegraphics[width=9cm]{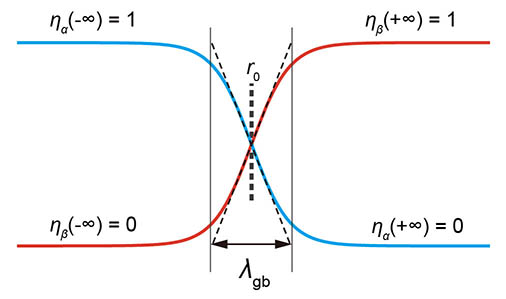}
	\caption{Schematic of the diffusive grain boundary and the profile of $\{\eta_\alpha\}$ across such interface.}
	\label{figA1}
\end{figure}

Considering the profiles of order parameters across the surface of a grain, where $\rho$ and only one $\eta$ varies from a semi-finite atmosphere/pore phase ($-\infty$) to a semi-finite grain phase ($+\infty$), as shown in Fig. \ref{figA1}. As for the normal direction $\mathbf{r}$ of an arbitrary point on the surface, profiles of $\rho$  and $\eta$ should satisfy the boundary conditions
\begin{equation}  
\left\{  
\begin{array}{lr}  
\rho=\eta=0  &  r \rightarrow -\infty  \\  
\rho=\eta=1  &  r \rightarrow +\infty  \\  
\nabla_\textbf{r}\rho = \nabla_\textbf{r}\eta = 0 & r \rightarrow \pm\infty 
\end{array}  
\right.
\label{eqA1}
\end{equation} 
Then the specific surface energy can be calculated by Cahn’s approach \citep{33cahn1961spinodal,38cahn1958free}
\begin{equation}
\gamma_\text{sf}=\int_{-\infty}^\infty \left[ f(T,\rho,\{\eta,0\}) + \frac{1}{2}T\kappa_\rho \left|\nabla_\mathbf{r}\rho \right|^2 +   \frac{1}{2}T\kappa_\eta \left|\nabla_\mathbf{r}\eta \right|^2 \right],
\label{eqA2}
\end{equation}
where the local free energy $f(T,\rho,\{\eta,0\})$ follows Eq. (\ref{eq16}). At equilibrium, functional in Eq. (\ref{eqA2}) maintains minimum at each temperature $T$, which requires $\rho$ and $\eta$ to satisfy the Euler-Lagrange equation
\begin{equation}
\frac{\delta\mathscr{F}(T,\rho,\{\eta,0\})}{\delta \rho} = \frac{\delta\mathscr{F}(T,\rho,\{\eta,0\})}{\delta \eta}=0.
\label{eqA3}
\end{equation}
According to the functional formulation in Eq. (\ref{eq3}), Eq. (\ref{eqA3}) can be rewritten in the following form
\begin{equation}
\frac{\partial f(T,\rho,\{\eta,0\})}{\partial\rho}-T\kappa_\rho\nabla^2_\mathbf{r}\rho = \frac{\partial f(T,\rho,\{\eta,0\})}{\partial\eta}-T\kappa_\eta\nabla^2_\mathbf{r}\eta = 0 .
\label{eqA4}
\end{equation}

In this model, the constraint of order parameters $(1-\rho)+\sum_{\alpha} \eta_\alpha=1$ should be held in any region within the substance at any time. Assuming $\rho$ and $\eta$ adopt the same shape as shown in Fig. \ref{figA1}, i.e. $\rho(\mathbf{r})=\eta(\mathbf{r})$, from Eq. (\ref{eqA4}) we can yield the following equation
\begin{equation}
\frac{1}{{{\kappa }_{\rho }}}\frac{\partial f\left( T,\rho ,\left\{ \eta ,0 \right\} \right)}{\partial \rho }=\frac{1}{{{\kappa }_{\eta }}}\frac{\partial f\left( T,\rho ,\left\{ \eta ,0 \right\} \right)}{\partial \eta },
\label{eqA5}
\end{equation}
where
\begin{equation*}
	\left\{ \begin{aligned}
		& \frac{\partial f\left( T,\rho ,\left\{ \eta ,0 \right\} \right)}{\partial \rho }=\underline{A}{{f}_{\text{ht}}}\left( T \right)+\underline{C}\left( 2\rho -6{{\rho }^{2}}+4{{\rho }^{3}} \right)+\underline{D}\left( 2\rho -6{{\eta }^{2}}+4{{\eta }^{3}} \right), \\ 
		& \frac{\partial f\left( T,\rho ,\left\{ \eta ,0 \right\} \right)}{\partial \eta }=\underline{B}{{f}_{\text{ht}}}\left( T \right)+\underline{D}\left[ 12\left( 1-\rho  \right)\eta -12\left( 2-\rho  \right){{\eta }^{2}}+12{{\eta }^{3}} \right]. \\ 
	\end{aligned} 
	\right.
\end{equation*}
Replacing every $\eta$ by $\rho$, we obtain:
\begin{equation}
\frac{1}{{{\kappa }_{\rho }}}\left[ \underline{A}{{f}_{\text{ht}}}\left( T \right)+\left( \underline{C}+\underline{D} \right)\left( 2\rho -1 \right)\left( 2\rho -2 \right)\rho \right]=\frac{1}{{{\kappa }_{\eta }}}\left[ \underline{B}{{f}_{\text{ht}}}\left( T \right)+\left( 6\underline{D} \right)\left( 2\rho -1 \right)\left( 2\rho -2 \right)\rho \right].
\label{eqA6}
\end{equation}

To make Eq. (\ref{eqA6}) hold at any $T$ and $\rho$, one can assume
\begin{subequations}
	\begin{equation}
	\left\{ \begin{aligned}
	& \frac{{\underline{A}}}{{{\kappa }_{\rho }}}=\frac{{\underline{B}}}{{{\kappa }_{\eta }}}, \\ 
	& \frac{\underline{C}+\underline{D}}{{{\kappa }_{\rho }}}=\frac{6\underline{D}}{{{\kappa }_{\eta }}}, \\ 
	\end{aligned} \right.
	\label{eqA7a}
	\end{equation}
	where
	\begin{equation*}
		\begin{aligned}
			& \underline{C}={{{\underline{C}}}_{\text{pt}}}-{{{\underline{C}}}_{\text{cf}}}\left( T-{{T}_{0}} \right), \\ 
			& \underline{D}={{{\underline{D}}}_{\text{pt}}}-{{{\underline{D}}}_{\text{cf}}}\left( T-{{T}_{0}} \right). \\ 
		\end{aligned}
	\end{equation*}
	Known that $\underline{A}+\underline{B}=1$, strong constraint above can be also rewritten as
	\begin{equation}
	\left\{ \begin{aligned}
	& \underline{A}=\frac{{{\kappa }_{\rho }}}{{{\kappa }_{\rho }}+{{\kappa }_{\eta }}},\text{      }\underline{B}=\frac{{{\kappa }_{\eta }}}{{{\kappa }_{\rho }}+{{\kappa }_{\eta }}}, \\ 
	& \frac{{{{\underline{C}}}_{\text{pt}}}+{{{\underline{D}}}_{\text{pt}}}}{{{\kappa }_{\rho }}}=\frac{6{{{\underline{D}}}_{\text{pt}}}}{{{\kappa }_{\eta }}},\frac{{{{\underline{C}}}_{\text{cf}}}+{{{\underline{D}}}_{\text{cf}}}}{{{\kappa }_{\rho }}}=\frac{6{{{\underline{D}}}_{\text{cf}}}}{{{\kappa }_{\eta }}}. \\ 
	\end{aligned} \right.
	\label{eqA7b}
	\end{equation}
\label{eqA7}
\end{subequations}
Eq. (\ref{eqA7b}) shows the relation between the model parameters $\underline{A},~\underline{B},~{{\underline{C}}_{\text{pt}}},~{{\underline{C}}_{\text{cf}}},~{{\underline{D}}_{\text{pt}}},~{{\underline{D}}_{\text{cf}}},~{{\kappa }_{\rho }}$ and ${{\kappa }_{\eta }}$. We can furthermore yield the expression ${{\gamma }_{\text{sf}}}$ for the specific surface energy by replacing $\rho$ by $\eta$ in Eq. (\ref{eqA2}). According to Eq. (\ref{eqA3}), one has
\begin{equation}
\frac{\partial f\left( T,\rho ,\left\{ \rho ,0 \right\} \right)}{\partial \rho }-T\left( {{\kappa }_{\rho }}+{{\kappa }_{\eta }} \right)\nabla _{\mathbf{r}}^{2}\rho =0.
\label{eqA8}
\end{equation}
By integrating both sides of Eq. (\ref{eqA8}), one can get \citep{10kuczynski1949self}
\begin{equation}
f\left( T,\rho ,\left\{ \eta ,0 \right\} \right)-\frac{1}{2}T\left( {{\kappa }_{\rho }}+{{\kappa }_{\eta }} \right){{\left| {{\nabla }_{\mathbf{r}}}\rho  \right|}^{2}}={{C}_{1}},
\label{eqA9}
\end{equation}
and the arbitrary constant ${{C}_{1}}$ equals zero considering the boundary conditions in the Eq. (\ref{eqA1}). Then one can obtain the explicit formulation of the specific surface energy at equilibrium by substituting Eq. (\ref{eqA9}) into Eq. (\ref{eqA2}), i.e.
\begin{equation}
\begin{split}
{{\gamma }_{\text{sf}}}&=\int_{-\infty }^{\infty }{\left[ 2f\left( T,\rho ,\left\{ \rho ,0 \right\} \right) \right]\text{d}r}, \\ 
&=\text{2}\sqrt{\frac{T\left( {{\kappa }_{\eta }}+{{\kappa }_{\rho }} \right)}{2}}\int_{0}^{1}{\sqrt{{{f}_{\text{ht}}}\rho +\left( \underline{C}+7\underline{D} \right){{\left( 1-\rho  \right)}^{2}}{{\rho }^{2}}}\text{d}\rho }. \\ 
\end{split}
\label{eqA10}
\end{equation}

Likewise, grain boundary energy can be derived. In this case, we consider two grains where the density variation is neglected. Then we have $\rho \approx 1$ across them. There are only two order parameters  ${{\eta }_{\alpha }}$ and  ${{\eta }_{\beta }}$  indicating the grains of both side. Assuming an isotropic grain boundary, profiles of ${{\eta }_{\alpha }}$ and ${{\eta }_{\beta }}$ across a grain boundary can be described by the following boundary conditions
\begin{equation}
\left\{  
\begin{array}{lr}  
{{\eta }_{\alpha }}=1,{{\eta }_{\beta }}=0  &  r\to -\infty  \\
{{\eta }_{\alpha }}=0,{{\eta }_{\beta }}=1  &  r\to +\infty  \\ 
{{\nabla }_{\mathbf{r}}}{{\eta }_{\alpha }}={{\nabla }_{\mathbf{r}}}{{\eta }_{\beta }}=0  &  r\to \pm \infty \\ 
\end{array}  
\right.
\label{eqA11}
\end{equation}
and the specific grain boundary energy can be calculated by
\begin{equation}
{{\gamma }_{\text{gb}}}=\int_{-\infty }^{\infty }{\left[ f\left( T,1,\left\{ {{\eta }_{\alpha }},{{\eta }_{\beta }} \right\} \right)+\frac{1}{2}T{{\kappa }_{\eta }}{{\left| {{\nabla }_{\mathbf{r}}}{{\eta }_{\alpha }} \right|}^{2}}+\frac{1}{2}T{{\kappa }_{\eta }}{{\left| {{\nabla }_{\mathbf{r}}}{{\eta }_{\beta }} \right|}^{2}} \right]}\text{d}r.
\label{eqA12}
\end{equation}
At equilibrium, the Euler-Lagrange equation across the grain boundary reads 
\begin{equation}
\frac{\delta\mathscr{F}\left( T,1,\left\{ {{\eta }_{\alpha }},{{\eta }_{\beta }} \right\} \right)}{\delta{{\eta }_{\alpha }}}=\frac{\delta\mathscr{F}\left( T,1,\left\{ {{\eta }_{\alpha }},{{\eta }_{\beta }} \right\} \right)}{\delta{{\eta }_{\beta }}}=0,
\label{eqA13}
\end{equation}
then
\begin{equation}
\frac{\partial f\left( T,1,\left\{ {{\eta }_{\alpha }},{{\eta }_{\beta }} \right\} \right)}{\partial {{\eta }_{\alpha }}}-T{{\kappa }_{\eta }}\nabla _{\mathbf{r}}^{2}{{\eta }_{\alpha }}=\frac{\partial f\left( T,1,\left\{ {{\eta }_{\alpha }},{{\eta }_{\beta }} \right\} \right)}{\partial {{\eta }_{\beta }}}-T{{\kappa }_{\eta }}\nabla _{\mathbf{r}}^{2}{{\eta }_{\beta }}=0 .
\label{eqA14}
\end{equation}
Since we constrain the summation of all non-conserved ${{\eta }_{\alpha }}$ to be unity, i.e. ${{\eta }_{\beta }}=1-{{\eta }_{\alpha }}$, it is easy to find that ${{\nabla }_{\mathbf{r}}}{{\eta }_{\alpha }}=-{{\nabla }_{\mathbf{r}}}{{\eta }_{\beta }}$. We can thereby take a single order parameter $\eta $ to replace ${{\eta }_{\alpha }}$ and ${{\eta }_{\beta }}$ in Eq. (\ref{eqA12}) by setting ${{\eta }_{\alpha }}=\eta $ and ${{\eta }_{\beta }}=1-\eta $. Following Eq. (\ref{eqA13}), it yields
\begin{equation}
\frac{\partial f\left( T,1,\left\{ \eta ,1-\eta  \right\} \right)}{\partial \eta }-2T{{\kappa }_{\eta }}\nabla _{\mathbf{r}}^{2}\eta =0.
\label{eqA15}
\end{equation}
By integrating both sides of Eq. (\ref{eqA15}) one can get
\begin{equation}
f\left( T,1,\left\{ {{\eta }_{\alpha }},{{\eta }_{\beta }} \right\} \right)-T{{\kappa }_{\eta }}{{\left| {{\nabla }_{\mathbf{r}}}\eta  \right|}^{2}}={{C}_{2}},
\label{eqA16}
\end{equation}
and the arbitrary constant ${{C}_{2}}$ equals zero considering the boundary conditions in  Eq. (\ref{eqA11}). Substituting Eq. (\ref{eqA16}) into Eq. (\ref{eqA12}), we eventually obtain the explicit formulation of the specific grain-boundary energy at equilibrium
\begin{equation}
\begin{split}
{{\gamma }_{\text{gb}}}&=\int_{-\infty }^{\infty }{\left[ 2f\left( T,1,\left\{ \eta ,1-\eta  \right\} \right) \right]\text{d}r}, \\ 
&=\text{2}\sqrt{T{{\kappa }_{\eta }}}\int_{0}^{1}{\sqrt{{{f}_{\text{ht}}}+12\underline{D}{{\left( 1-\eta  \right)}^{2}}{{\eta }^{2}}}\text{d}\eta }. 
\end{split}
\label{eqA17}
\end{equation}

We regard Eq. (\ref{eqA10}) and (\ref{eqA17}) as the relations which link the model parameters $\underline{A}$, $\underline{B}$, $\ {{\underline{C}}_{\text{pt}}}$, $\ {{\underline{C}}_{\text{cf}}}$, $\ {{\underline{D}}_{\text{pt}}}$, ${{\underline{D}}_{\text{cf}}}$, ${{\kappa }_{\eta }}$ and ${{\kappa }_{\rho }}$ to the measurable, temperature-dependent material properties ${{\gamma }_{\text{sf}}}\left( T \right)$ and ${{\gamma }_{\text{gb}}}\left( T \right)$, which are usually linearly fitted in practical measurement. Due to the dependency, there are only four independent quantities in these eight model parameters, i.e. ${{\underline{D}}_{\text{pt}}}\text{, }{{\underline{D}}_{\text{cf}}},{{\kappa }_{\rho }},{{\kappa }_{\eta }}$ (or ${{\underline{C}}_{\text{pt}}}\text{, }{{\underline{C}}_{\text{cf}}},{{\kappa }_{\rho }},{{\kappa }_{\eta }}$). Two methods can be utilized to obtain those parameters:

1) \textit{Direct method.} This method is to determine them by directly formulating 4 equations based on Eqs. (\ref{eqA10}) and (\ref{eqA17}) with the experimental values of ${{\gamma }_{\text{sf}}}$ and ${{\gamma }_{\text{gb}}}$ at two different temperature. However, solving those equations is difficult since the integration contains the undetermined quantities $\underline{C}$ and $\underline{D}$.

2) \textit{Indirect method.} This method firstly relate the ${{\kappa }_{\rho }},{{\kappa }_{\eta }}$ and ${{\underline{C}}_{\text{pt}}}$, ${{\underline{D}}_{\text{pt}}}$ to the experimental values of the surface and grain-boundary energies at $T_0$, i.e.
\begin{equation}
\begin{split}
\gamma _{\text{sf}}^{{{T}_{0}}} &=\text{2}\sqrt{\frac{{{T}_{0}}\left( {{\kappa }_{\eta }}+{{\kappa }_{\rho }} \right)}{2}}\int_{0}^{1}{\sqrt{\left( {{{\underline{C}}}_{\text{pt}}}+7{{{\underline{D}}}_{\text{pt}}} \right){{\left( 1-\rho  \right)}^{2}}{{\rho }^{2}}}\text{d}\rho } \\ 
&=\frac{1}{3\sqrt{2}}\sqrt{{{T}_{0}}\left( {{\kappa }_{\eta }}+{{\kappa }_{\rho }} \right)\left( {{{\underline{C}}}_{\text{pt}}}+7{{{\underline{D}}}_{\text{pt}}} \right)}, \\ 
\gamma _{\text{gb}}^{{{T}_{0}}}&=\text{2}\sqrt{{{T}_{0}}{{\kappa }_{\eta }}}\int_{0}^{1}{\sqrt{12{{{\underline{D}}}_{\text{pt}}}{{\left( 1-\eta  \right)}^{2}}{{\eta }^{2}}}\text{d}\eta } \\ 
&=\frac{2}{\sqrt{3}}\sqrt{{{T}_{0}}{{\kappa }_{\eta }}{{{\underline{D}}}_{\text{pt}}}}, \\ 
\end{split}
\label{eqA18}
\end{equation}
where ${{\underline{C}}_{\text{pt}}}$ and ${{\underline{D}}_{\text{pt}}}$ are related according to Eq. (\ref{eqA7b}). In the phase-field model, the grain boundary is a diffusive interface, and its approximate width ${{\lambda }_{\text{gb}}}$ at $T_{0}$ can be expressed as
\begin{equation}
{{\lambda}^{T_{0}}_{\text{gb}}}\approx \frac{1}{\tan \left( \theta^{T_{0}} /2 \right)}=\frac{1}{{{\left. {{\nabla }_{\mathbf{r}}}\eta^{T_{0}}  \right|}_{{{r}_{0}}}}} = \sqrt{4{{T}_{0}}{{\kappa }_{\eta }}/3{{{\underline{D}}}_{\text{pt}}}} .
\label{eqA19}
\end{equation}
Based on Eq. (\ref{eqA7b}), at $T_0$ we can define a another fitting parameter
\begin{equation}
\nu ={{T}_{0}}{{\underline{D}}_{\text{cf}}}/{{\underline{D}}_{\text{pt}}}={{T}_{0}}{{\underline{C}}_{\text{cf}}}/{{\underline{C}}_{\text{pt}}}
\label{eqA19-1}
\end{equation}
By using the four Eqs. (\ref{eqA18})--(\ref{eqA19-1}) at $T_0$, we can determine $\underline{D}_\text{pt}$, $\underline{D}_\text{cf}$, $\kappa_\rho$, $\kappa_\eta$ and thus eight model parameters according to the dependency. Then according to Eqs. (\ref{eqA10}) and (\ref{eqA17}), the temperature-dependent $\gamma_\text{sf} \left(T \right)$ and $\gamma_\text{gb}\left( T \right)$ can be obtained. By tuning ${\lambda }_{\text{gb}}^{T_0}$ and $\nu$, we can fit ${{\gamma }_{\text{sf}}}\left( T \right)$ and ${{\gamma }_{\text{gb}}}\left( T \right)$ to make them close to the experimental data, as shown in Fig. \ref{fig11}.

\section{Further formulation for the finite element implementation}\label{appendixc}
In this appendix we present the detailed formulations for the finite element implementation within the MOOSE framework. Transient solver based on backward Euler algorithm is employed. We firstly introduce the residuals for the weak form in Eq. (\ref{eq27}), i.e. 
\begin{equation}
\left\{ \begin{aligned}
& {{R}_{\mu }}=\int_{\Omega }{{\dot{\rho }}}{{\psi }_{\mu }}\text{d}\Omega +\int_{\Omega }{{{M}_{ij}}{{\mu }_{,j}}{{\psi }_{\mu ,i}}}\text{d}\Omega -\int_{\Gamma }{{{M}_{ij}}{{\mu }_{,j}}}{{\psi }_{\mu }}{{n}_{i}} \text{d}\Gamma \\ 
& {{R}_{\rho }}=\int_{\Omega }{\frac{\partial f}{\partial \rho }}{{\psi }_{\rho }}\text{d}\Omega -\int_{\Omega }{\mu {{\psi }_{\rho }}}\text{d}\Omega - \int_{\Omega }{{{\kappa }_{\rho }}T{{\rho }_{,i}}{{\psi }_{\rho ,i}}}\text{d}\Omega + \int_{\Gamma }{{{\kappa }_{\rho }}T{{\rho }_{,i}}}{{\psi }_{\rho }}{{n}_{i}}\text{d}\Gamma \\ 
& {{R}_{{{\eta }_{\alpha }}}}=\int_{\Omega }{{{{\dot{\eta }}}_{\alpha }}}{{\psi }_{\eta }}\text{d}\Omega + L\int_{\Omega }{\frac{\partial f}{\partial {{\eta }_{\alpha }}}{{\psi }_{\eta }}}\text{d}\Omega +L\int_{\Omega }{{{\kappa }_{\eta }}T{{\eta }_{\alpha ,i}}}{{\psi }_{\eta ,i}}\text{d}\Omega -L\int_{\Gamma }{{{\kappa }_{\eta }}T{{\eta }_{\alpha ,i}}}{{\psi }_{\eta }}{{n}_{i}}\text{d}\Gamma \\ 
& {{R}_{T}}=\int_{\Omega }{\frac{\partial e}{\partial T}\dot{T}}{{\psi }_{T}}\text{d}\Omega +\int_{\Omega }{\frac{\partial e}{\partial \rho }\dot{\rho }{{\psi }_{T}}}\text{d}\Omega +\sum\limits_{\alpha }{\int_{\Omega }{\frac{\partial e}{\partial {{\eta }_{\alpha }}}}{{{\dot{\eta }}}_{\alpha }}{{\psi }_{T}}\text{d}\Omega -}\int_{\Omega }{k{{T}_{,i}}}{{\psi }_{T,i}}\text{d}\Omega +\int_{\Gamma }{k{{T}_{,i}}}{{\psi }_{T}}{{n}_{i}}\text{d}\Gamma \\ 
\end{aligned} \right.
\label{eqA30}
\end{equation}
Following the Galerkin approach, the test functions are discretized as
\begin{equation*}
{{\psi }_{\mu }}=\varphi _{\mu }^{I}\psi _{\mu }^{I} ~ , ~ {{\psi }_{\rho }}=\varphi _{\rho }^{I}\psi _{\rho }^{I} ~ , ~ {{\psi }_{\eta }}=\varphi _{\eta }^{I}\psi _{\eta }^{I} ~ , ~ {{\psi }_{T}}=\varphi _{T}^{I}\psi _{T}^{I}
\end{equation*}
where $I$ denotes the node index and Einstein summation convention is used; $\psi _{\mu }^{I}$, $\psi _{\rho }^{I}$, $\psi _{\eta_\alpha }^{I}$ and $\psi _{T}^{I}$ are the nodal weights while $\varphi _{\mu}^{I}$, $\varphi _{\rho}^{I}$, $\varphi _{\eta_\alpha }^{I}$ and $\varphi _{T}^{I}$ are the shape functions for the corresponding variables. Then we can find the discretized residuals from Eq. (\ref{eqA30}) as
\begin{equation}
\left\{ \begin{aligned}
& R_{\mu }^{I}=\int_{\Omega }{{\dot{\rho }}}\varphi _{\mu }^{I}\text{d}\Omega +\int_{\Omega }{{{M}_{ij}}{{\mu }_{,j}}\varphi _{\mu ,i}^{I}}\text{d}\Omega -\int_{\Gamma }{{{M}_{ij}}{{\mu }_{,j}}}\varphi _{\mu }^{I}{{n}_{i}}\text{d}\Gamma \\ 
& R_{\rho }^{I}=\int_{\Omega }{\frac{\partial f}{\partial \rho }}\varphi _{\rho }^{I}\text{d}\Omega -\int_{\Omega }{\mu \varphi _{\rho }^{I}}\text{d}\Omega -\int_{\Omega }{{{\kappa }_{\rho }}T{{\rho }_{,i}}}\varphi _{\rho ,i}^{I}\text{d}\Omega +\int_{\Gamma }{{{\kappa }_{\rho }}T}{{\rho }_{,i}}\varphi _{\rho }^{I}{{n}_{i}}\text{d}\Gamma \\ 
& R_{{{\eta }_{\alpha }}}^{I}=\int_{\Omega }{{{{\dot{\eta }}}_{\alpha }}}\varphi _{\eta }^{I}\text{d}\Omega +L\int_{\Omega }{\frac{\partial f}{\partial {{\eta }_{\alpha }}}\varphi _{\eta }^{I}}\text{d}\Omega +L\int_{\Omega }{{{\kappa }_{\eta }}T{{\eta }_{\alpha ,i}}}\varphi _{\eta ,i}^{I}\text{d}\Omega -L\int_{\Gamma }{{{\kappa }_{\eta }}T{{\eta }_{\alpha ,i}}}\varphi _{\eta }^{I}{{n}_{i}}\text{d}\Gamma \\ 
& R_{T}^{I}=\int_{\Omega }{\frac{\partial e}{\partial T}\dot{T}}\varphi _{T}^{I}\text{d}\Omega +\int_{\Omega }{\frac{\partial e}{\partial \rho }\dot{\rho }\varphi _{T}^{I}}\text{d}\Omega +\sum\limits_{\alpha }{\int_{\Omega }{\frac{\partial e}{\partial {{\eta }_{\alpha }}}}{{{\dot{\eta }}}_{\alpha }}\varphi _{T}^{I}\text{d}\Omega -}\int_{\Omega }{k{{T}_{,i}}\varphi _{T,i}^{I}\text{d}\Omega } +\int_{\Gamma }{k{{T}_{,i}}}\varphi _{T}^{I}{{n}_{i}}\text{d}\Gamma \\ 
\end{aligned} \right.
\label{eqa31}
\end{equation}
Similarly, the variable fields $\mu $, $\rho $, ${{\eta }_{\alpha }}$, $T$ as well as their time derivative $\dot{\mu }$, $\dot{\rho }$, ${{\dot{\eta }}_{\alpha }}$, $\dot{T}$ are also discretized by the shape functions, i.e.
\begin{equation*}
\begin{aligned}
&\rho =\varphi _{\rho }^{I}{{\rho }^{I}} ~ , ~ \mu =\varphi _{\mu }^{I}{{\mu }^{I}} ~ , ~ {{\eta }_{\alpha }}=\varphi _{\eta }^{I}{{\eta }_{\alpha }}^{I} ~ , ~ T=\varphi _{T}^{I}{{T}^{I}}; \\
&\dot{\rho }=\varphi _{\rho }^{I}{{\dot{\rho }}^{I}} ~ , ~ \dot{\mu }=\varphi _{\mu }^{I}{{\dot{\mu }}^{I}} ~ , ~ {{\dot{\eta }}_{\alpha }}=\varphi _{\eta }^{I}{{\dot{\eta }}_{\alpha }}^{I} ~ , ~ \dot{T}=\varphi _{T}^{I}{{\dot{T}}^{I}},
\end{aligned}
\end{equation*}
where ${{\mu }^{I}}$, ${{\rho }^{I}}$, $\eta _{\alpha }^{I}$ and ${{T}^{I}}$ are the nodal values. The linearization of the residuals gives the following element-level linear equations
\begin{equation}
\begin{split}
\left[ \begin{matrix}
R_{\mu }^{I}  \\
R_{\rho }^{I}  \\
R_{{{\eta }_{1}}}^{I}  \\
\vdots   \\
R_{{{\eta }_{N}}}^{I}  \\
R_{T}^{I}  \\
\end{matrix} \right] =&-\left[ \begin{matrix}
K_{\mu ,\mu }^{IJ} & K_{\mu ,\rho }^{IJ} & K_{\mu ,{{\eta }_{1}}}^{IJ} & \cdots  & K_{\mu ,{{\eta }_{N}}}^{IJ} & K_{\mu ,T}^{IJ}  \\
K_{\rho ,\mu }^{IJ} & K_{\rho ,\rho }^{IJ} & K_{\rho ,{{\eta }_{1}}}^{IJ} & \cdots  & K_{\rho ,{{\eta }_{N}}}^{IJ} & K_{\rho ,T}^{IJ}  \\
K_{{{\eta }_{1}},\mu }^{IJ} & K_{{{\eta }_{1}},\rho }^{IJ} & K_{{{\eta }_{1}},{{\eta }_{1}}}^{IJ} & \cdots  & K_{{{\eta }_{1}},{{\eta }_{N}}}^{IJ} & K_{{{\eta }_{1}},T}^{IJ}  \\
\vdots  & \vdots  & \vdots  & \ddots  & \vdots  & \vdots   \\
K_{{{\eta }_{N}},\mu }^{IJ} & K_{{{\eta }_{N}},\rho }^{IJ} & K_{{{\eta }_{N}},{{\eta }_{1}}}^{IJ} & \cdots  & K_{{{\eta }_{N}},{{\eta }_{N}}}^{IJ} & K_{{{\eta }_{N}},T}^{IJ}  \\
K_{T,\mu }^{IJ} & K_{T,\rho }^{IJ} & K_{T,{{\eta }_{1}}}^{IJ} & \cdots  & K_{T,{{\eta }_{N}}}^{IJ} & K_{T,T}^{IJ}  \\
\end{matrix} \right]\left[ \begin{matrix}
\delta {{\mu }^{J}}  \\
\delta {{\rho }^{J}}  \\
\delta \eta _{1}^{J}  \\
\vdots   \\
\delta \eta _{N}^{J}  \\
\delta {{T}^{J}}  \\
\end{matrix} \right] \\ 
 & - \left[\begin{matrix}
D_{\mu ,\mu }^{IJ} & D_{\mu ,\rho }^{IJ} & D_{\mu ,{{\eta }_{1}}}^{IJ} & \cdots  & D_{\mu ,{{\eta }_{N}}}^{IJ} & D_{\mu ,T}^{IJ}  \\
D_{\rho ,\mu }^{IJ} & D_{\rho ,\rho }^{IJ} & D_{\rho ,{{\eta }_{1}}}^{IJ} & \cdots  & D_{\rho ,{{\eta }_{N}}}^{IJ} & D_{\rho ,T}^{IJ}  \\
D_{{{\eta }_{1}},\mu }^{IJ} & D_{{{\eta }_{1}},\rho }^{IJ} & D_{{{\eta }_{1}},{{\eta }_{1}}}^{IJ} & \cdots  & D_{{{\eta }_{1}},{{\eta }_{N}}}^{IJ} & D_{{{\eta }_{1}},T}^{IJ}  \\
\vdots  & \vdots  & \vdots  & \ddots  & \vdots  & \vdots   \\
D_{{{\eta }_{N}},\mu }^{IJ} & D_{{{\eta }_{N}},\rho }^{IJ} & D_{{{\eta }_{N}},{{\eta }_{1}}}^{IJ} & \cdots  & D_{{{\eta }_{N}},{{\eta }_{N}}}^{IJ} & D_{{{\eta }_{N}},T}^{IJ}  \\
D_{T,\mu }^{IJ} & D_{T,\rho }^{IJ} & D_{T,{{\eta }_{1}}}^{IJ} & \cdots  & D_{T,{{\eta }_{N}}}^{IJ} & D_{T,T}^{IJ}  \\
\end{matrix} \right]\left[ \begin{matrix}
\delta {{{\dot{\mu }}}^{J}}  \\
\delta {{{\dot{\rho }}}^{J}}  \\
\delta \dot{\eta }_{1}^{J}  \\
\vdots   \\
\delta \dot{\eta }_{N}^{J}  \\
\delta {{{\dot{T}}}^{J}}  \\
\end{matrix} \right].
\end{split}
\label{eqa32}
\end{equation}
The terms in the tangent matrix and the damping matrix are calculated by $K_{\xi ,\zeta }^{IJ}={\partial R_{\xi }^{I}} / {\partial {{\zeta }^{J}}}$ and $D_{\xi ,\zeta }^{IJ}={\partial R_{\xi }^{I}} / {\partial {{{\dot{\zeta }}}^{J}}}$, which give the non-zero terms as follows

\begin{equation}
	\begin{aligned}
	& K_{\mu ,\mu }^{IJ}=\int_{\Omega }{{{M}_{ij}}\varphi _{\mu ,j}^{J}\varphi _{\mu ,i}^{I}}\text{d}\Omega - \int_{S}{{{M}_{ij}}\varphi _{\mu ,j}^{J}}\varphi _{\mu }^{I}{{n}_{i}}\text{d}\Gamma,
	\\
	& K_{\mu ,\rho }^{IJ}=\int_{\Omega }{\frac{\partial {{M}_{ij}}}{\partial \rho }\varphi _{\rho }^{J}}\mu_{,j}\varphi _{\mu ,i}^{I}\text{d}\Omega -\int_{\Gamma }{\frac{\partial {{M}_{ij}}}{\partial \rho }\varphi _{\rho }^{J}} \mu_{,j} \varphi _{\mu }^{I}{{n}_{i}}\text{d}\Gamma, 	
	\\
	& K_{\mu ,{{\eta }_{\alpha }}}^{IJ}=\int_{\Omega }{\frac{\partial {{M}_{ij}}}{\partial {{\eta }_{\alpha }}}\varphi _{\eta }^{J}} \mu_{,j} \varphi _{\mu ,i}^{I}\text{d}\Omega -\int_{\Gamma } \frac{\partial {{M}_{ij}}}{\partial {{\eta }_{\alpha }}}\varphi _{\eta }^{J} \mu_{,j} \varphi _{\mu }^{I}{{n}_{i}}\text{d}\Gamma,
	\\
	& K_{\mu ,T}^{IJ}=\int_{\Omega } \frac{\partial {{M}_{ij}}}{\partial T}\varphi _{T}^{J} \mu_{,j}
	\varphi _{\mu ,i}^{I}\text{d}\Omega -\int_{\Gamma }{\frac{\partial {{M}_{ij}}}{\partial T}\varphi _{T}^{J}} \mu_{,j} \varphi _{\mu }^{I}{{n}_{i}}\text{d}\Gamma,
	\\ 
	& K_{\rho ,\mu }^{IJ}=-\int_{\Gamma }{\varphi _{\mu }^{J}\varphi _{\rho }^{I}}\text{d}\Gamma ,
	\\  
	&K_{\rho ,\rho }^{IJ}=\int_{\Omega }{\frac{{{\partial }^{2}}f}{\partial {{\rho }^{2}}}}\varphi _{\rho }^{J}\varphi _{\rho }^{I}\text{d}\Omega - \int_{\Omega }{{{\kappa }_{\rho }} T \varphi _{\rho ,i}^{J}}\varphi _{\rho ,i}^{I}\text{d}\Omega + \int_{\Gamma }{{{\kappa }_{\rho }} T \varphi _{\rho ,i}^{J}}\varphi _{\rho }^{I}{{n}_{i}}\text{d}\Gamma ,
	\\
	&K_{\rho ,{{\eta }_{\alpha }}}^{IJ}=\int_{\Omega }{\frac{{{\partial }^{2}}f}{\partial \rho \partial {{\eta }_{\alpha }}}}\varphi _{\eta }^{J}\varphi _{\rho }^{I}\text{d}\Omega ,
	\\
	&K_{\rho ,T}^{IJ}=\int_{\Omega }{\frac{{{\partial }^{2}}f}{\partial \rho \partial T}}\varphi _{T}^{J}\varphi _{\rho }^{I}\text{d}\Omega - \int_{\Omega }{{{\kappa }_{\rho }}\varphi _{T}^{J} \rho_{,i}  \varphi_{\rho ,i}^{I}} \text{d}\Omega + \int_{\Gamma }{{\kappa }_{\rho }}\varphi _{T}^{J} \rho_{,i} \varphi _{\rho }^{I}{{n}_{i}}\text{d}\Gamma ,
	\\
	&K_{{{\eta }_{\alpha }},\rho }^{IJ}=L\int_{\Omega }{\frac{{{\partial }^{2}}f}{\partial \rho \partial \mu }\varphi _{\rho }^{J}\varphi _{\eta }^{I}}\text{d}\Omega ,
	\\
	&K_{{{\eta }_{\alpha }},{{\eta }_{\alpha }}}^{IJ}=L\int_{\Omega }{\frac{{{\partial }^{2}}f}{\partial {{\eta }_{\alpha }}^{2}}\varphi _{\eta }^{J}\varphi _{\eta }^{I}} \text{d}\Omega +L\int_{\Omega }{{{\kappa }_{\eta }} T \varphi _{\eta ,i}^{J}}\varphi _{\eta ,i}^{I}\text{d}\Omega -L\int_{\Gamma }{{{\kappa }_{\eta }} T \varphi _{\eta ,i}^{J}}\varphi _{\eta }^{I}{{n}_{i}}\text{d}\Gamma ,
	\\
	&K_{{{\eta }_{\alpha }},T}^{IJ}=L\int_{\Omega }{\frac{\partial f}{\partial {{\eta }_{\alpha }}\partial T}\varphi _{T}^{J}} \varphi _{\eta }^{I} \text{d}\Omega +L\int_{\Omega }{{\kappa }_{\eta }}\varphi _{T}^{J} \eta_{\alpha ,i} \varphi_{\eta ,i}^{I}\text{d}\Omega -L\int_{S}{{\kappa }_{\eta }}\varphi _{T}^{J} \eta_{\alpha ,i} \varphi _{\eta }^{I}{{n}_{\Gamma }}\text{d}\Gamma ,
	\\ 
	&\begin{split}
	K_{T,\rho }^{IJ}=&\int_{\Omega }{\frac{{{\partial }^{2}}e}{\partial T\partial \rho }\varphi _{\rho }^{J}\dot{T}}\varphi _{T}^{I}\text{d}\Omega +\int_{\Omega }{\frac{{{\partial }^{2}}e}{\partial {{\rho }^{2}}}\varphi _{\rho }^{J}\dot{\rho }\varphi _{T}^{I}}\text{d}\Omega  \\ 
	&+\sum\limits_{\alpha } \int_{\Omega }{\frac{{{\partial }^{2}}e}{\partial {{\eta }_{\alpha }}\partial \rho }}\varphi _{\rho }^{J}{{{\dot{\eta }}}_{\alpha }}\varphi _{T}^{I}\text{d}\Omega - \int_{\Omega }{\frac{\partial k}{\partial \rho }\varphi _{\rho }^{J} T_{,i} \varphi_{T,i}^{I}\text{d}\Omega }+\int_{\Gamma }\frac{\partial k}{\partial \rho }\varphi _{\rho }^{J} T_{,i} \varphi _{T}^{I}{{n}_{i}}\text{d}\Gamma,  \\ 
	\end{split}
	\\ 
	&\begin{split}
	K_{T,{{\eta }_{\alpha }}}^{IJ}= &\int_{\Omega }{\frac{{{\partial }^{2}}e}{\partial T\partial {{\eta }_{\alpha }}}\varphi _{\eta }^{J}\dot{T}}\varphi_{T}^{I}\text{d}\Omega +\int_{\Omega }{\frac{{{\partial }^{2}}e}{\partial \rho \partial {{\eta }_{\alpha }}}\varphi _{\rho }^{J}\dot{\rho }\varphi _{T}^{I}}\text{d}\Omega  \\ 
	&+\int_{\Omega }{\frac{{{\partial }^{2}}e}{\partial {{\eta }_{\alpha }}^{2}}}\varphi _{\eta }^{J}{{{\dot{\eta }}}_{\alpha }}\varphi _{T}^{I}\text{d}\Omega - \int_{\Omega }{\frac{\partial k}{\partial {{\eta }_{\alpha }}}\varphi _{\eta }^{J}\varphi _{T,i}^{J}{{T}^{J}}\varphi _{T,i}^{I}\text{d}\Omega }+\int_{\Gamma }\frac{\partial k}{\partial {{\eta }_{\alpha }}} 
	\varphi _{\eta }^{J} T_{,i}  \varphi_{T}^{I}{{n}_{i}}\text{d}\Gamma,  \\ 
	\end{split}
	\\ 
	&\begin{split}
	K_{T,T}^{IJ} = &\int_{\Omega }{\frac{{{\partial }^{2}}e}{\partial {{T}^{2}}}\varphi _{T}^{J}\dot{T}}\varphi _{T}^{I}\text{d}\Omega +\int_{\Omega }{\frac{{{\partial }^{2}}e}{\partial \rho \partial T}\varphi _{\rho}^{J}\dot{\rho }\varphi _{T}^{I}}\text{d}\Omega  \\ 
	&+\sum\limits_{\alpha }{\int_{\Omega }{\frac{{{\partial }^{2}}e}{\partial {{\eta }_{\alpha }}\partial T}}\varphi _{\eta}^{J}{{{\dot{\eta }}}_{\alpha }}\varphi _{T}^{I}\text{d}\Omega +}\int_{\Omega }{k\varphi _{T,i}^{J}\varphi _{T,i}^{I}\text{d}\Omega +\int_{\Omega }{\frac{\partial k}{\partial T}\varphi_{T}^{J} T_{,i} \varphi _{T,i}^{I}\text{d}\Omega }} \\ 
	&+\int_{\Gamma }{k\varphi _{T,i}^{J}}\varphi_{T}^{I}{{n}_{i}}\text{d} \Gamma + \int_{\Gamma } \frac{\partial k}{\partial T}\varphi_{T}^{J} T_{,i} \varphi_{T}^{I}{{n}_{i}}\text{d}\Gamma,  \\ 
	\end{split}
		\end{aligned}
	\end{equation}
and
\begin{equation}
\begin{aligned}
	&D_{\mu ,\rho }^{IJ}=\int_{\Omega }{\varphi _{\rho }^{J}\varphi _{\mu }^{I}}\text{d}\Omega ~ , ~
	D_{{{\eta }_{\alpha }},{{\eta }_{\alpha }}}^{IJ}=\int_{\Omega }{\varphi _{{{\eta }_{\alpha }}}^{J}\varphi _{{{\eta }_{\alpha }}}^{I}}\text{d}\Omega  \\
	&D_{T,\rho }^{IJ}=\int_{\Omega }{\frac{\partial e}{\partial \rho }\varphi _{\rho }^{J}\varphi _{T}^{I}}\text{d}\Omega ~ , ~
	D_{T,\eta_\alpha }^{IJ}=\int_{\Omega }{\frac{\partial e}{\partial \eta_\alpha }\varphi _{\eta }^{J}\varphi _{T}^{I}}\text{d}\Omega ~ , ~
	D_{T,T}^{IJ}=\int_{\Omega }{\frac{\partial e}{\partial T}\varphi _{T}^{J}\varphi _{T}^{I}}\text{d}\Omega.
	\end{aligned}
\end{equation}




\section*{References}
\bibliography{reference}

\end{document}